\documentclass[12pt,a4paper]{article}

\usepackage{epsfig}
\usepackage{psfig}
\usepackage{exscale}
\def\lsim{\raise0.3ex\hbox{$<$\kern-0.75em\raise-1.1ex\hbox{$\sim$}}}
\def\gsim{\raise0.3ex\hbox{$>$\kern-0.75em\raise-1.1ex\hbox{$\sim$}}}
\newcommand{\be}{\begin{equation}}
\newcommand{\ee}{\end{equation}}
\newcommand{\ba}{\begin{eqnarray}}
\newcommand{\ea}{\end{eqnarray}}

\def\spose#1{\hbox to 0pt{#1\hss}}
\def\ltapprox{\mathrel{\spose{\lower 3pt\hbox{$\mathchar"218$}}
 \raise 2.0pt\hbox{$\mathchar"13C$}}}
\def\gtapprox{\mathrel{\spose{\lower 3pt\hbox{$\mathchar"218$}}
 \raise 2.0pt\hbox{$\mathchar"13E$}}}
\def\PRep{{ Phys.\ Rept.\ }}

\def\NT{{N_{\tau}}}
\def\nt{\ifmmode\NT\else$\NT$\fi}
\def\NS{N_\sigma}
\def\ns{\ifmmode\NS\else$\NS$\fi}

\def\p{^\prime}
\def\v{\vec}
\def\n{\noindent}

\def\PBP{{\langle \bar\psi\psi \rangle }}
\def\phv{\v\phi}
\def\vH{\v H}
\begin{document}
\begin{titlepage} 
\thispagestyle{empty}

 \mbox{} \hfill BI-TP 2004/40\\
 \mbox{} \hfill May 2005\\
% \mbox{} \hfill hep-lat/??
\begin{center}
\vspace*{0.8cm}
{{\Large \bf Scaling and Goldstone effects in a QCD\\
with two flavours of adjoint quarks\\
}}\vspace*{1.0cm}
{\large J. Engels, S. Holtmann and T. Schulze}\\ \vspace*{0.8cm}
\centerline {{\em Fakult\"at f\"ur Physik, 
    Universit\"at Bielefeld, D-33615 Bielefeld, Germany}} \vspace*{0.4cm}
\protect\date \\ \vspace*{0.9cm}
{\bf   Abstract   \\ } \end{center} \indent
We study QCD with two Dirac fermions in the adjoint representation at
finite temperature by Monte Carlo simulations. In such a theory the 
deconfinement and chiral phase transitions occur at different temperatures.
We locate the second order chiral transition point at $\beta_c=5.624(2)$
and show that the scaling behaviour of the chiral condensate in the vicinity
of $\beta_c$ is in full agreeement with that of the $3d$ $O(2)$ universality
class, and to a smaller extent comparable to the  $3d$ $O(6)$ class. From
the previously determined first order deconfinement transition point 
$\beta_d=5.236(3)$ and the two-loop beta function we
find the ratio $T_c/T_d\approx 7.8(2)$. In the region between the two phase
transitions we explicitly confirm the quark mass dependence of the chiral
condensate which is expected due to the existence of Goldstone modes like
in $3d$ $O(N)$ spin models. At the deconfinement transition $\PBP$ shows
a gap, and below $\beta_d$, it is nearly mass-independent for fixed $\beta$.

\vfill \begin{flushleft} 
\noindent{\rule[-.3cm]{5cm}{.02cm}} \\
\vspace*{0.2cm} 
E-mail: engels, holtmann, tschulze@physik.uni-bielefeld.de\
\end{flushleft} 
\end{titlepage}

%%%%%%%%%%%%%%%%%%%%%%%%%%%%%%%%%%%%%%%%%%%%%%%%%%%%%%%%%%%%%%%%%%%%%%%%%%%%%%%%

\section{Introduction}
\label{section:Intro}

%%%%%%%%%%%%%%%%%%%%%%%%%%%%%%%%%%%%%%%%%%%%%%%%%%%%%%%%%%%%%%%%%%%%%%%%%%%%%%%%
The classification of the chiral transition in finite temperature quantum
chromodynamics (QCD) for two flavours of quarks is a long-standing problem.
This refers in particular to analyses of lattice data obtained from simulations
with staggered fermions. For the continuum theory a second order chiral 
transition is predicted to belong to the universality class of the 
three-dimensional $O(4)$ spin model \cite{Pisarski:1984ms,Wilczek:1992sf, 
Rajagopal:1992qz}. 
Due to the lattice formulation of the fermionic sector of
the theory, chiral symmetry is lost for Wilson fermions and reduced
to an $O(2)$ symmetry for staggered fermions. Full symmetry is then
recovered only in the continuum limit. Surprisingly, corresponding tests
with Wilson fermions \cite{Iwasaki:1996ya,AliKhan:2000iz} confirmed the 
proposed universality class, whereas the first tests with staggered
fermions and the standard QCD action remained unsatisfactory 
\cite{Karsch:1993tv}-\cite{Bernard:1999fv}. Essentially, the comparison
of universal critical 
quantities from $O(N)$ spin models to lattice QCD data failed in the 
close vicinity of the chiral transition. However, at temperatures above the
transition some agreement was found on the pseudocritical line, namely for 
the form of the line itself and for finite size scaling of the chiral 
condensate on the line \cite{Engels:2001bq}. The apparent failure of the 
tests was attributed to a very small critical region, i.\ e.\ use of too
large $m_qa$-values and lattice artifacts, i.\ e.\ calculations with too
small $\NT$ and/or unimproved actions. 

The issue is still under consideration and still unsettled:
Kogut and Sinclair, who use a QCD action including an irrelevant 4-fermion 
term, conclude from simulations at $m_q=0$ on a $24^3\times 8$ lattice
\cite{Kogut:2004ia} that the critical region must be very small, but the 
transition class could be compatible to the $3d$ $O(2)/O(4)$ universality
class. Chandrasekharan and Strouthos \cite{Chandrasekharan:2004kd}
performed simulations at strong coupling and compared their data 
to predictions from chiral perturbation theory for $O(2)$. They
find full agreement for masses which are much smaller than those normally
used in QCD calculations. Contrary to that,
D' Elia et al.\ \cite{D'Elia:2005bv} exclude $O(2)/O(4)$ scaling in their
finite-size-scaling analysis of specific heat and susceptibility data for 
the standard QCD action. Instead they find consistency of their data with
a first order transition.

The difficulties and discrepancies encountered in the investigation of 
the nature of the chiral transition could be related to another puzzle of
finite temperature QCD: the coincidence of the chiral and the deconfinement
transitions. Though there is no a priori reason why the two transitions 
should occur at the same temperature in QCD, they seem to be strongly 
bound to each other. The interplay of the deconfinement and chiral
transitions could then obscure the critical chiral behaviour in the 
vicinity of the common critical point. It is possible to study
the properties of the chiral transition without direct interference from
the deconfinement transition. In a special QCD-related model, $SU(N_c)$ 
gauge theory with two Dirac fermions in the adjoint representation (aQCD),
the deconfinement transition occurs at a temperature $T_d$ which is
definitely smaller than $T_c$, the chiral transition temperature 
\cite{Kogut:1985xa,Karsch:1998qj}. 

In general, aQCD with $N_f$ massless fermions exhibits an $SU(2N_f)$ 
chiral symmetry, which is spontaneously broken at low 
temperatures to $SO(2N_f)$ \cite{Peskin:1980gc,Smilga:1994tb}.
Since, in contrast to QCD, the fermionic part of the aQCD action does not
break the global $Z(N_c)$ centre symmetry of the $SU(N_c)$ gauge part,
the model holds in addition the same global $Z(N_c)$ symmetry as pure
$SU(N_c)$ gauge theory. Correspondingly, one expects two finite
temperature transitions and there are two independent order parameters,
the Polyakov loop and the chiral condensate, which indicate the breaking
of the respective symmetries. The universality class of the chiral 
transition of aQCD is normally different from that of QCD.
For two flavours of adjoint Dirac fermions the chiral symmetry group is
$SU(4)$. That group is isomorphic to $O(6)$, which breaks to $O(5)$,
implying the existence of 5 Goldstone bosons. The breaking 
$SU(4)\rightarrow SO(4)$ leads however to 9 Goldstone bosons. The 
corresponding transition belongs therefore to a different universality
class. In a recent paper \cite{Basile:2004wa} this issue has been 
investigated using renormalization-group arguments and first results
for the critical exponents of this class have been obtained: 
$\eta\approx 0.2\, ,\nu\approx 1.1$. The points we have just raised pertain 
to continuum theory. On the lattice, the staggered formulation
leaves, like in QCD, only a remaining $O(2)$ chiral symmetry. In the
continuum limit one should however recover the full $SU(4)$ symmetry
and the critical behaviour following from its breaking to $SO(4)$.

The main objective of this paper is to investigate the chiral transition
of aQCD and if possible to establish its universality class. Since we are 
simulating aQCD on the lattice with two flavours of staggered Dirac     
fermions\footnote{We have adopted the QCD convention of counting the 
number of fermion flavours. Because the fermions are in the adjoint
representation one Dirac fermion corresponds to two Majorana fermions.} 
we expect to find the three-dimensional $O(2)$ class. The 
thermodynamics of aQCD with $N_c=3$ has been studied to a great extent  
already by Karsch and L\"utgemeier \cite{Karsch:1998qj}. In particular,
a first order deconfinement transition was found at $\beta_d=5.236(3)$  
and a second order chiral transition at $\beta_c=5.79(5)$ was inferred
from the mass dependence of the chiral condensate and its susceptibility.
However, no comparison to the universal scaling functions of $O(N)$
spin models was attempted in the vicinity of $\beta_c$. In order to 
achieve that we have made additional simulations at more $\beta$-values
around the supposed chiral transition point and at additional and smaller 
$m_qa$-values. A second aim of our paper is the study of the Goldstone 
effects which are connected to the breaking of chiral symmetry. If aQCD 
behaves as an effective three-dimensional $O(N)$ spin model close to the
chiral transition point then one would expect the chiral susceptibility
$\chi_m$ to diverge below $\beta_c$ as $(m_qa)^{-1/2}$ for 
$m_qa\rightarrow 0$. Such a mass dependence has in fact been observed
in the range $\beta_d< \beta < \beta_c$ in Ref.\ \cite{Karsch:1998qj}.
For smaller $T$-values, approaching $T=0$, the theory changes to an 
effective four-dimensional theory and the expected divergence of $\chi_m$ 
becomes logarithmic in $m_qa$. The change in the behaviours of $\PBP$
and $\chi_m$ with decreasing $T$, also below $\beta_d$, is therefore of
interest. 

The paper is organized as follows. In the next section we define the 
action and the observables of aQCD, which we want to consider. Then we
discuss numerical details of our simulations and give an overview of the
available data. In Section \ref{section:O(N)} we recapitulate those 
results from the $O(N)$ spin models which are most relevant for the
subsequent analysis of the aQCD data in Section \ref{section:analysis}.
We close with a summary and the conclusions. 

%%%%%%%%%%%%%%%%%%%%%%%%%%%%%%%%%%%%%%%%%%%%%%%%%%%%%%%%%%%%%%%%%%%%%%%%%%%%%%%%

\section{QCD with adjoint fermions on the lattice}
\label{section:Aqcd}

%%%%%%%%%%%%%%%%%%%%%%%%%%%%%%%%%%%%%%%%%%%%%%%%%%%%%%%%%%%%%%%%%%%%%%%%%%%%%%%%

In the following we shall use previous results obtained by Karsch and 
L\"utgemeier and extend their simulations for our purposes. The version
of aQCD which we use is therefore the same as the one already described in
detail in Ref.\ \cite{Karsch:1998qj} and it will suffice to address only 
the main aspects of the corresponding lattice theory. The action is similar
to QCD with staggered fermions in the fundamental representation. The gluon
part $S_G$ of the action is the standard Wilson one-plaquette gauge 
action with $SU(3)$ link variables $U^{(3)}$ and coupling $\beta = 6/g^2$
\be 
S_G = \beta \sum_P \{ 1- {1 \over 3}{\rm Re~tr}\,U^{(3)}_P \}~.
\label{sglue} 
\ee 
In contrast to QCD the fermions are in the eight-dimensional adjoint 
representation of the $SU(3)$ group and have eight colour degrees of 
freedom. The fermionic part $S_F$ of the action requires then also an 
eight-dimensional representation $U^{(8)}$ of the link matrices which may
be expressed in terms of the three-dimensional matrices 
\be
U^{(8)}_{ab} ={1 \over 2} {\rm tr}[\lambda_a U^{(3)} \lambda_b 
U^{(3)\dagger}]~,
\label{ueight}
\ee
where the $\lambda_a$ are the eight Gell-Mann matrices and tr is as in 
(\ref{sglue}) the 3-trace. The total aQCD action is 
\be
S=S_G +S_F = S_G + \sum_{x,y}  \bar\psi_x M(U^{(8)})_{x,y} \psi_y~. 
\label{action}
\ee  
Here, $M(U^{(8)})$ is the standard staggered fermion matrix where the
link variables $U^{(3)}$ have been replaced by the respective $U^{(8)}$
matrices. We note some important properties of $U^{(8)}$. From Eq.\ 
(\ref{ueight}) it follows that $U^{(8)}$ is real and its 8-trace 
(${\rm tr}_8$) is related to the 3-trace of $U^{(3)}$ via
\be
{\rm tr}_8\, U^{(8)} = |{\rm tr}\,U^{(3)} |^2 -1~.
\label{trace}
\ee  
Further, $U^{(8)}$ is invariant under $Z(3)$ transformations of the 
$U^{(3)}$-matrices because the centre elements appear in Eq.\ 
(\ref{ueight}) as a complex conjugate pair. As a consequence, $S_F$
does not break the $Z(3)$ centre symmetry and the Polyakov loop
\be
L_3 = \left \langle\: \frac{1}{\ns^3}\left |\sum_{\v x}{\rm tr}\,
\prod_{x_4=1}^{\nt }U^{(3)}(x_4,\v x)\right |\: \right \rangle ~,
\label{polya}   
\ee 
is an order parameter for the deconfinement transition like in pure 
$SU(3)$ gauge theory in the limit $\ns \rightarrow \infty$. 
The order parameter for the chiral transition, the chiral condensate,
is defined by  
\be
\PBP = {T\over V}{\partial \ln Z \over \partial m_qa} = {1 \over \ns^3\nt}
{N_f \over 2} \langle\: {\rm Tr} M^{-1}\: \rangle~.
\label{chicon}
\ee
where the trace Tr has to be taken over the eight colours and all
spacepoints. The derivative of the chiral condensate with respect to 
$m_qa$ is the chiral susceptibility
\be
\chi_m ={\partial \PBP \over \partial m_qa} = 
{T\over V}{\partial^2 \ln Z \over \partial(m_qa)^2}~.
\label{chisus}
\ee
It is the sum of the connected term $\chi_{con}$ and the disconnected
term $\chi_{dis}$. The latter measures the fluctuations of the chiral
condensate. The two parts are given by
\ba
\chi_{dis}\!\!&=&\!\!{1 \over N_{\sigma}^3 N_{\tau}}\left({N_f \over 2}
\right)^2 \left( \langle\: ({\rm Tr}M^{-1})^2 \:\rangle - 
\langle\: {\rm Tr}M^{-1}\:\rangle ^2  \right)~,\label{dis} \\
\chi_{con}\!\!&=&\!\! - {N_f \over 2} \sum_x \langle\: M_{x,0}^{-1}
M_{0,x}^{-1} \:\rangle~. \label{con}
\ea
In our subsequent analysis we shall utilise the data for the Polykov loop, 
the chiral condensate and the disconnected part of the susceptibility.

Since the matrices $U^{(8)}$ are real this is also the case for the total 
fermion matrix $M(U^{(8)})$. The determinant of $M(U^{(8)})$ may then be 
represented as a Gaussian integral over real bosonic fields with the 
advantage that the exact Hybrid Monte Carlo algorithm can be used for the 
simulation of aQCD with two adjoint fermion flavours. The corresponding
effective pseudo-fermion action is given by
\be
S_{eff} = S_G(U^{(3)}) + \Phi^t \left( M(U^{(8)})^t M(U^{(8)})\right)\Phi~,
\label{seff}
\ee
with real eight-component $\Phi$-fields, which reside only on even lattice
sites to compensate for the fermion doubling introduced by the use of 
$M^tM$ instead of $M$. The molecular dynamics equations for aQCD are 
more complicated as the ones for QCD, because $M$ is now quadratic in
$U^{(3)}$. Details of the update algorithm can be found in appendix A 
of Ref.\ \cite{L"utgemeier:1998}. 

Our simulations were performed on $\ns^3\times 4$ lattices with 
$\ns = 8,12$ and 16 and a fixed length $\tau=n \delta\tau=0.25\,$ of
the trajectories. The number of molecular dynamics steps $n$ was chosen
such that the final Metropolis acceptance rate was in the range of
$60-80\%$. On the $8^3\times 4$ lattice $n$ varied between 20 and 40
for $0.10\ge m_qa\ge 0.02$. For lower masses and/or increasing spatial 
extension $\ns a$ of the lattice the number of steps had to be increased.
The optimal $n$-value was however relatively independent of $\beta$,
apart from the region around and below the deconfinement transition,
where larger $n$ were necessary. The chiral condensate and the
Polyakov loop were measured

%----------------------------------------------------------------------------
\setlength{\unitlength}{1cm}
\begin{picture}(10,19)
\put(1.8,14){
   \epsfig{bbllx=127,bblly=265,bburx=451,bbury=588,
       file=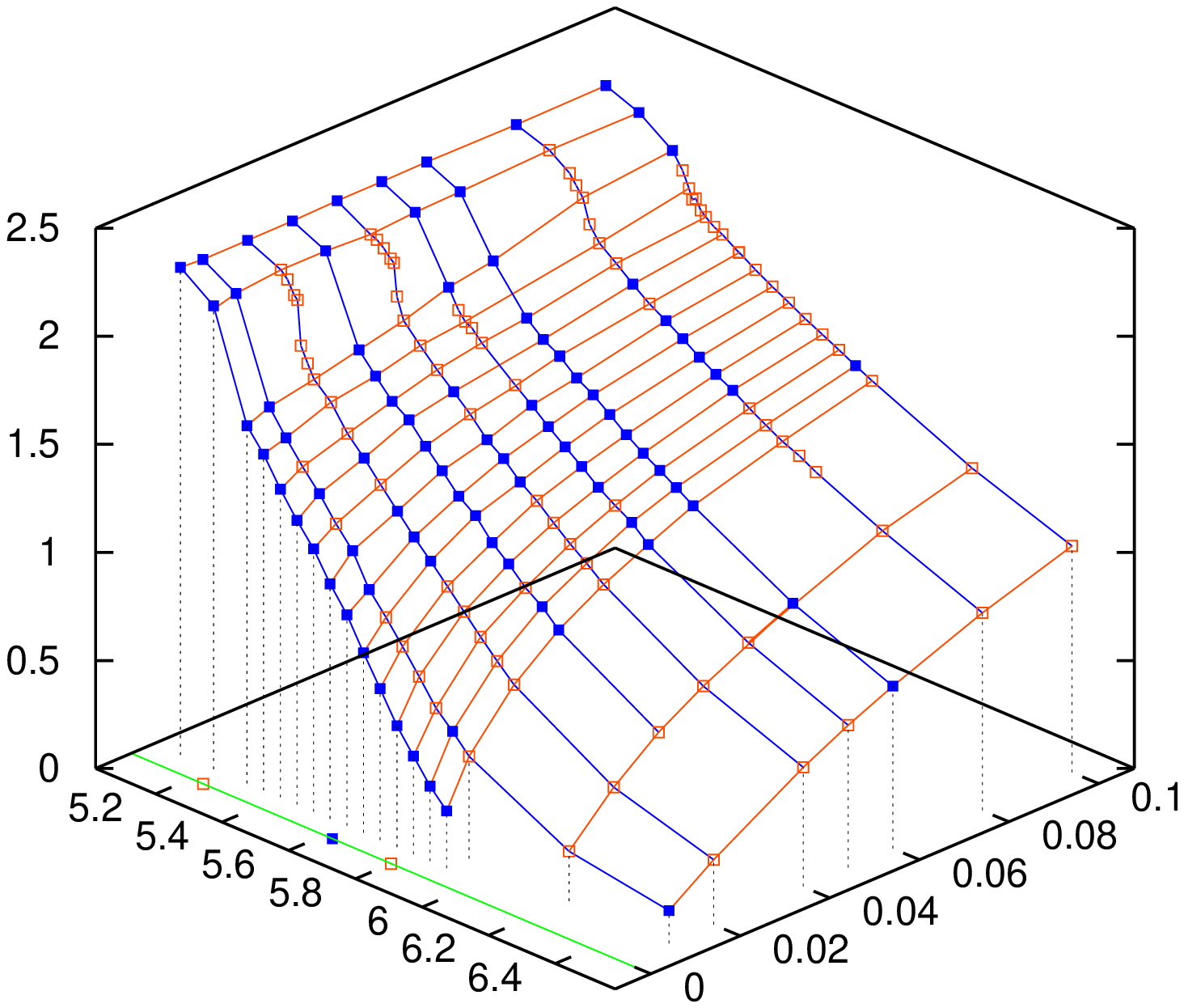, width=90mm}
          }  
\put(1.8,4){ 
   \epsfig{bbllx=127,bblly=265,bburx=451,bbury=588,
       file=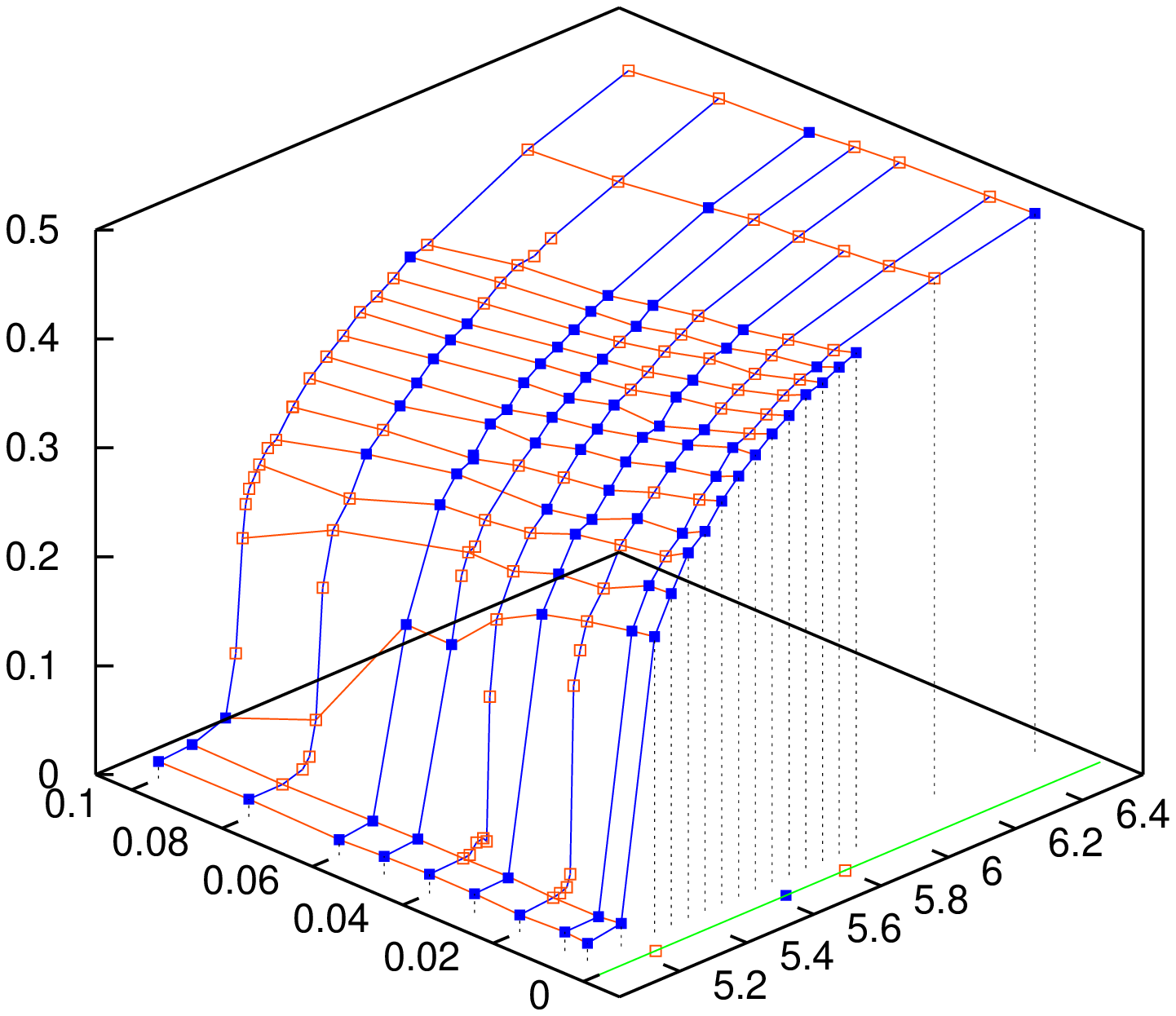, width=90mm}
          }
\put(1.3,18){\large{$\PBP$}}
\put(2.8,10){\large{$\beta$}}
\put(10.3,10){\large{$m_q a$}}
\put(12.5,14.3){\large{(a)}}
\put(1.5,8){\large{$L_3$}}
\put(10.5,0){\large{$\beta$}}
\put(2.4,0){\large{$m_q a$}}
\put(12.5,4.3){\large{(b)}}
\end{picture}
%----------------------------------------------------------------------------
%------------------------------------------------------------------------
\begin{figure}[b!]
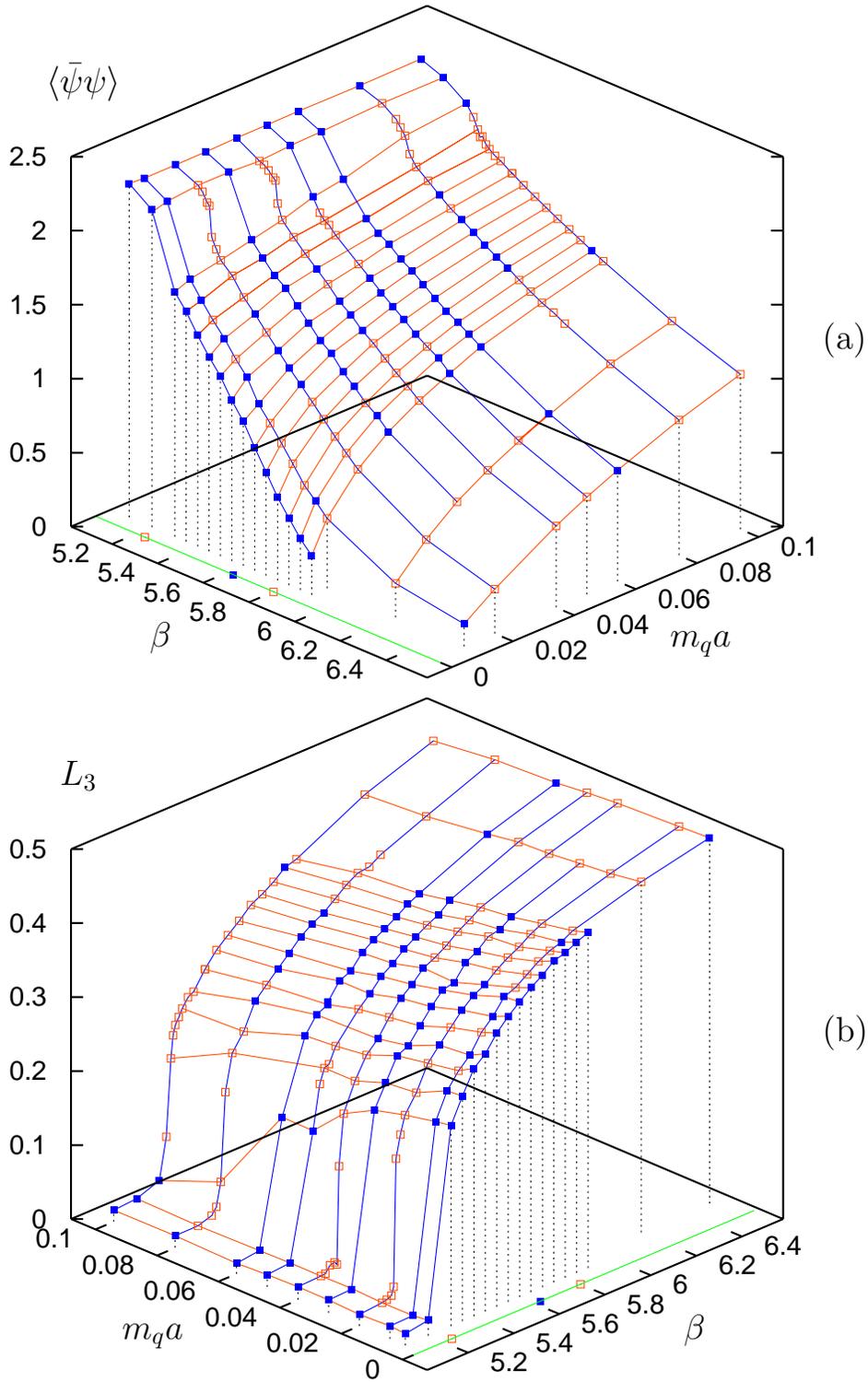

\caption{The chiral condensate $\PBP$ (a) and the Polyakov loop $L_3$
(b) versus $\beta$ and $m_qa$ on the $8^3\times 4$ lattice (plot (b) has
been rotated by $90^{\circ}$). The filled squares denote the new data, the
empty ones are from Ref.\ \cite{Karsch:1998qj}. The dotted lines indicate the 
height. The data points are connected by straight lines to guide the eye.}
\label{fig:3dplot}
\end{figure}

\newpage
\n after the completion of each trajectory. Both the condensate and
the disconnected part of the susceptibility were calculated from
a noisy estimator \cite{Bitar:1989dn} with 25 random source vectors.
As already mentioned in the introduction, our first intention was
to enlarge the data set of Ref.\ \cite{Karsch:1998qj}. To this end we
have performed additional simulations on the $8^3\times 4$ lattice
with 900-2000 trajectories for measurements at 
$m_qa = 0.005, 0.03$ and 0.06, and for the new 
$\beta$-values 5.1, 5.45, 5.55, 5.6 and $5.65\,$. The already existing 
data at $m_qa= 0.01,\,0.02,\,0.04,\,0.05,\,0.08$ and 0.10 and at the
$\beta$-values 5.2, 5.3, 5.35, 5.4, 5.5, 5.7, 5.75, 5.85, 5.9, 6.2 and
6.5 have been completed to obtain a mesh covering the region from above
the chiral phase transition to a point just below the deconfinement
transition. In order to see possible finite size dependencies we have
carried out further simulations at selected points on larger lattices
with $\ns=12$ and 16. The respective data will be show in the figures
of Section 4. The chiral condensate data from the different lattices
coincide inside the error bars at essentially all points. In the chiral 
susceptibility data there are however some differences which may 
be due to insufficient statistics. 

An overview of the data from the $8^3\times 4$ lattice is given in 
Fig.\ \ref{fig:3dplot} where we show three-dimensional plots of the chiral
condensate $\PBP$ and the Polyakov loop $L_3$. In the plots we have
indicated on the line at $m_qa=0$ the previously \cite{Karsch:1998qj}
determined transition points $\beta_d=5.236(3)$ and $\beta_c=5.79(5)$
by empty squares. The filled square on the same line denotes our new
estimate for $\beta_c$, which we will explain in Section 4. Obviously, 
the chiral phase transition point cannot be determined by a simple
inspection of the order parameter $\PBP$, even though we have many new 
data points in the critical region. For the first order deconfinement
transition the situation is much clearer. The Polyakov loop $L_3$
shows jumps between $\beta=5.2$ and $5.3\,$, depending on the mass value. 
Our new results at $\beta=5.1$ and 5.2 are definitely below the
deconfinement transition. Insofar they support the determination of 
the exact position by Karsch and L\"utgemeier who produced more data in 
the transition region at the masses $m_qa = 0.02,0.04,0.08$ and $0.10\,$.
The straight line connections in Fig.\ \ref{fig:3dplot} in this region
at the other mass values should therefore be taken with care. Two further 
comments are appropriate here. Already in Ref.\ \cite{L"utgemeier:1998} 
a remarkable fact was noted: the first order deconfinement transition
point $\beta_d=5.236(3)$ of aQCD coincides with the second order chiral 
transition point $\beta_c= 5.233(7)$ of QCD with two fundamental quarks 
and the standard staggered action \cite{Laermann:1998}. In addition, 
we see in Fig.\ \ref{fig:3dplot} that $\PBP$ exhibits also a 
discontinuity at $\beta_d\,$, whereas the Polyakov loop shows no sign
of the chiral transition.  

%%%%%%%%%%%%%%%%%%%%%%%%%%%%%%%%%%%%%%%%%%%%%%%%%%%%%%%%%%%%%%%%%%%%%%%%%%%%%%%%

\section{The three-dimensional $O(N)$ spin models}
\label{section:O(N)}

%%%%%%%%%%%%%%%%%%%%%%%%%%%%%%%%%%%%%%%%%%%%%%%%%%%%%%%%%%%%%%%%%%%%%%%%%%%%%%%%

In this section we want to summarize the basic facts about the 
three-dimensional $O(N)$ spin models and their universality classes.
A more general survey on $O(N)$-symmetric systems can be found in Ref.\
\cite{Pelissetto:2000ek}. The numerical results to which we will refer 
in the following have been obtained from simulations of the standard
$O(N)$-invariant non-linear sigma models with the partition function 
\be 
Z = \int [ d\phv ] \exp \left[ J \sum\limits_{x,y} \phv_x \cdot
\phv_y + \vH \cdot \sum\limits_x \phv_x \right]~. 
\label{pfct}
\ee 
Here, $x$ and $y$ are nearest-neighbour sites on a three-dimensional 
hypercubic lattice and $\phv_x$ is an $N$-component unit vector at site 
$x$. The coupling $J$ acts as inverse temperature $J=1/T$ and $\vH$
is the magnetic field. The spin vector $\phv_x$ may be decomposed into a
longitudinal (parallel to $\vH$) and a transverse component
(perpendicular to $\vH$): $\phv_x = \phi_x^{\Vert} \hat{\vH} +
\phv_x^{\bot}\,$. The order parameter of the system, the magnetization
$M$ and its susceptibility $\chi_L$ are then defined by
\be
 M=\left< {1\over V}\sum\limits_x {\phi}_x^{\Vert}\right> =
\left< {\phi}^{\Vert}\right>, \quad \quad 
 \chi_L = {\partial M \over \partial H} = V \left(\left< 
{\phi}^{\Vert 2}\right> -M^2\right)~,
\label{Oobsev}
\ee
that is, $M$ is the expectation value of the lattice average of the
longitudinal spin component and $\chi_L$ its derivative with respect 
to $H=|\vH|$ (implying $H\ge 0$) .% The volume is $V=L^3$ and $L$
%is the number of lattice points per direction 

The spin system exhibits a second order phase transition at a critical 
temperature $T_c=1/J_c$. For $H\rightarrow 0$ and $T<T_c$ the
$O(N)$-symmetry of the spin system is spontaneously broken and the 
magnetization attains a finite value $M(T,0)>0$, whereas $M(T>T_c,0)=0$.
The behaviour of $M(T,H)$ in the region around the critical point is
described by the equation of state 
\be
M = h^{1/\delta} f_G (z)~, \quad {\rm with} \quad z= t/h^{1/\beta\delta}~,
\label{eos}
\ee  
where $f_G(z)$ is a universal scaling function, which is normalized
such that
\be
f_G(0)\; =\; 1~,\quad {\rm and}\quad f_G(z) {\raisebox{-1ex}{$
\stackrel{\displaystyle\longrightarrow}{\scriptstyle z \rightarrow -\infty}$}}
(-z)^{\beta}~.
\label{normfg}
\ee
This is achieved by the introduction of the reduced temperature $t$ and
field $h$ with
\be
t= (T-T_c)/T_0 \quad  {\rm and}\quad h=H/H_0~,
\label{redu}
\ee  
such that
\be
 M(t=0,h) = h^{1/\delta}~,\quad M (t<0,H \rightarrow 0) =(-t)^{\beta}~.
\label{critb}
\ee
The critical exponents $\delta$ and $\beta$ determine all other critical
exponents via the hyperscaling relations (here $d=3$)
\be
\quad \gamma \;=\; \beta (\delta -1), \quad d\nu \;=\; \beta (1 +\delta) 
\;=\; 2-\alpha, \quad2-\eta  \;=\; \gamma/\nu~.
\label{hypsc}
\ee
The scaling function $f_G(z)$ and the critical exponents are the same for
all members of the corresponding universality class. They can therefore be
used in tests on the class of a second order phase transition. In a series 
of papers these functions have been calculated for the three-dimensional
$O(N)$ models with $N=2$ \cite{Engels:2000xw}, $N=4$ 
\cite{Toussaint:1996qr,Engels:1999wf,Engels:2003nq} and $N=6$
\cite{Holtmann:2003he}.
%------------------------------------------------------------------------
\begin{figure}[t!]
\begin{center}
   \epsfig{bbllx=127,bblly=265,bburx=451,bbury=588,
       file=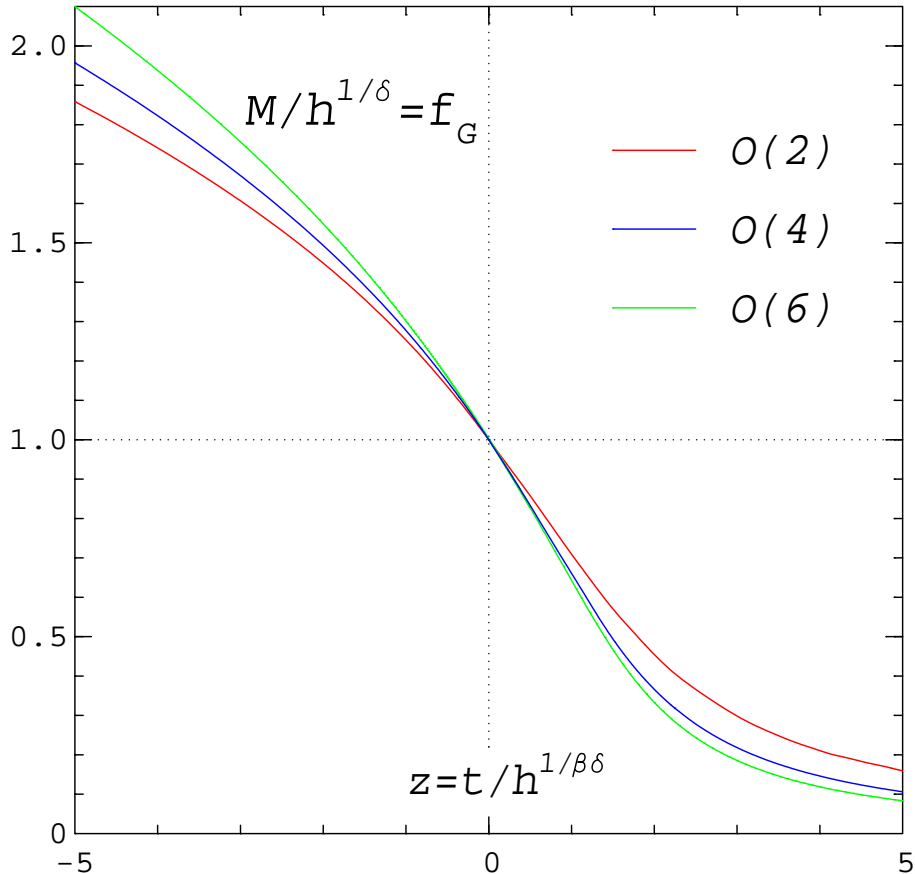, width=110mm}
\end{center}
\vspace*{0.2cm}
\caption{The universal scaling function $f_G=Mh^{-1/\delta}$ versus the
scaling variable $z=t h^{-1/\beta\delta}$ for $N=2$ \cite{Engels:2000xw},
$N=4$ \cite{Engels:2003nq} and $N=6$ \cite{Holtmann:2003he}. 
Parametrizations of the curves can be found in Refs.\ \cite{Engels:2000xw,
Engels:2003nq,Holtmann:2003he}.}
\label{fig:fgon}
\end{figure}
%------------------------------------------------------------------------
They are shown in Fig.\ \ref{fig:fgon}. We observe similar shapes of the
different $f_G(z)$-curves with an increasing slope for larger $N$.
In the plot the critical line $T=T_c$ is mapped to the point $z=0$, the
symmetric (high temperature) phase $T>T_c, H\rightarrow 0$ corresponds
to the limit $z \rightarrow \infty$ and the coexistence line $T<T_c,
H\rightarrow 0 $ to $z \rightarrow -\infty$. Since the exponents $\delta$ 
and $\beta$ are slightly different for each $N$ the scales of both axes
of the plot vary with $N$. In Table \ref{tab:O(N)} we have listed the
critical exponents which have been used in Fig.\ \ref{fig:fgon}.  
We see that the exponent $\beta$
%------------------------------------------------------------------------
\begin{table}[b]
\begin{center}
\begin{tabular}{|c||l|l|l|l||l|l||c|}
\hline
 $N$ & $~~\beta$ & $~~~\delta$ & $~~~\nu$ & $~~~\alpha$ 
 & $~1/\delta$ & $~1/\beta\delta$ & Ref.\ \\
\hline
 2 & 0.349 & 4.7798 & 0.6724 & -0.017 & 0.2092 & 0.5995 &
 \cite{Engels:2000xw} \\
 4 & 0.380 & 4.824  & 0.7377 & -0.213 & 0.2073 & 0.5455 & 
\cite{Engels:2003nq} \\
 6 & 0.425 & 4.780  & 0.818  & -0.454 & 0.2092 & 0.4922 &  
\cite{Holtmann:2003he} \\
\hline
\end{tabular}
\end{center}
\vspace{-0.2cm}
\caption{The critical exponents of the $3d$ $O(N)$ spin models.}
\label{tab:O(N)}
\end{table}
%------------------------------------------------------------------------

\n is increasing with $N$, while $\delta$ is nearly constant. The main 
influence of $N$ in a scaling plot is therefore coming from the 
$N$-dependence of the product $\Delta=\beta\delta$ which appears in the
scaling variable $z$.
 
Since the susceptibility $\chi_L$ is the derivative of $M$ with respect to
$H$, it can as well be expressed in terms of a universal scaling function
\be
\chi_L = {\partial M \over \partial H} = {h^{1/\delta -1} \over H_0}
 f_{\chi}(z)~,
\label{cscale}
\ee
with 
\be
f_{\chi}(z) = {1 \over \delta} \left( f_G(z) - {z \over \beta} f_G\p (z)
\right)~.
\label{fchi}
\ee
The scaling function $f_{\chi}(z)$ has a peak at a positive value
$z=z_{pc}$. Correspondingly, $\chi_L$ has a maximum for each fixed value
$h$ at a temperature $t_{pc}(h)$. This function of $h$ defines the 
pseudocritical line in the $(t,h)$-plane for $t\ge 0$. Obviously, it
is given by
\be
t_{pc}h^{-1/\Delta} = z_{pc} \quad \mbox{or}\quad T_{pc}-T_c =
c H^{1/\Delta}~,
\label{pseudo}
\ee
where $c=z_{pc}T_0H_0^{-1/\Delta}$ is a non-universal constant. Eq.\
(\ref{pseudo}) is often used to determine $T_c$. 

Due to the existence of massless Goldstone modes in $O(N)$ models with
$N>1$ and dimension $2<d\le 4$ singularities are expected on the whole 
coexistence line $T<T_c,H= 0$ \cite{Zinn-Justin:1996,Wallace:1975,
Anishetty:1995kj} in addition to the known critical behaviour at $T_c\,$.
In particular, the susceptibility $\chi_L$ is diverging on the 
coexistence line. For all $T<T_c$ and $H\rightarrow 0$ the divergence
is
\be
\chi_L (T,H) \sim \left\{ \begin{array}{l@{\quad {\rm for~}\,d=\,}l}
H^{-1/2} & 3\\ \ln H& 4 \end{array} \right.~,
\label{chidiv}
\ee
which implies the following $H$-dependence at fixed $T$ for the
magnetization
\be
M (T,H) =M(T,0)+c_1(T)\cdot \left\{ \begin{array}{l@{\quad {\rm for~}\,d=}l}
H^{1/2} & 3\\ H(\ln H-1)& 4 \end{array} \right.~.
\label{mdiv}
\ee
For the three-dimensional $O(N)$ models with $N=2,4,6$ the 
$H^{1/2}$-dependence of the magnetization for $T<T_c$ has been verified 
numerically in Refs.\ \cite{Engels:2000xw,Engels:1999wf,Holtmann:2003he}. 
In all cases one finds that the slope $c_1(T)$ is small for very low 
temperatures, that is $M$ is essentially constant. Approaching $T_c$, the 
slope increases and the region close to $H=0$ where $M$ is a straight line 
as a function of  $H^{1/2}$ shrinks. This is of course what one
expects, because at $T_c$ the magnetization changes its $H$-dependence from 
$H^{1/2}$ to $H^{1/\delta}$ with the effect that $c_1 \rightarrow \infty$.
The scaling functions $f_G$ and $f_{\chi}$ describe this behaviour
correctly in the critical region. Yet, Eqs.\ (\ref{chidiv}) and
(\ref{mdiv}) are also valid for all temperatures below the critical region. 
\clearpage

%------------------------------------------------------------------------
%%%%%%%%%%%%%%%%%%%%%%%%%%%%%%%%%%%%%%%%%%%%%%%%%%%%%%%%%%%%%%%%%%%%%%%%%%%%%%%%

\section{Analysis of the aQCD data}
\label{section:analysis}

%%%%%%%%%%%%%%%%%%%%%%%%%%%%%%%%%%%%%%%%%%%%%%%%%%%%%%%%%%%%%%%%%%%%%%%%%%%%%%%%

\n  In aQCD the temperature is given by $T=1/(\nt a)$ and the analogon
of the magnetic field is the quark mass $m_qa$. Since the lattice 
spacing $a$ is a monotone function of the coupling $\beta$, we can use 
$\beta$ instead of the temperature and temperature differences are 
proportional to coupling differences. Accordingly, we define instead of
the reduced temperature $t$ a reduced coupling $\beta_r$ and a reduced
field $m$, where
\be
\beta_r = (\beta -\beta_c) /\beta_0~,~ m=(m_qa)/(m_qa)_0 \quad 
\mbox{and} \quad z=\beta_r m^{-1/\Delta}~.
\label{redth}
\ee
The constants $\beta_0$ and $(m_qa)_0$ have to be fixed 
from the normalization of the scaling function for $\PBP$.  
It is clear, that before any scaling tests can be carried out, the
critical point $\beta_c$ of the chiral transition has to be determined 
as accurately as possible.
%%%%%%%%%%%%%%%%%%%%%%%%%%%%%%%%%%%%%%%%%%%%%%%%%%%%%%%%%%%%%%%%%%%%%%%%%%%%%%%%

\subsection{Location of the chiral transition point}
\label{sub:critpoint}

\begin{figure}[t]
\begin{center}
   \epsfig{bbllx=127,bblly=265,bburx=451,bbury=588,
       file=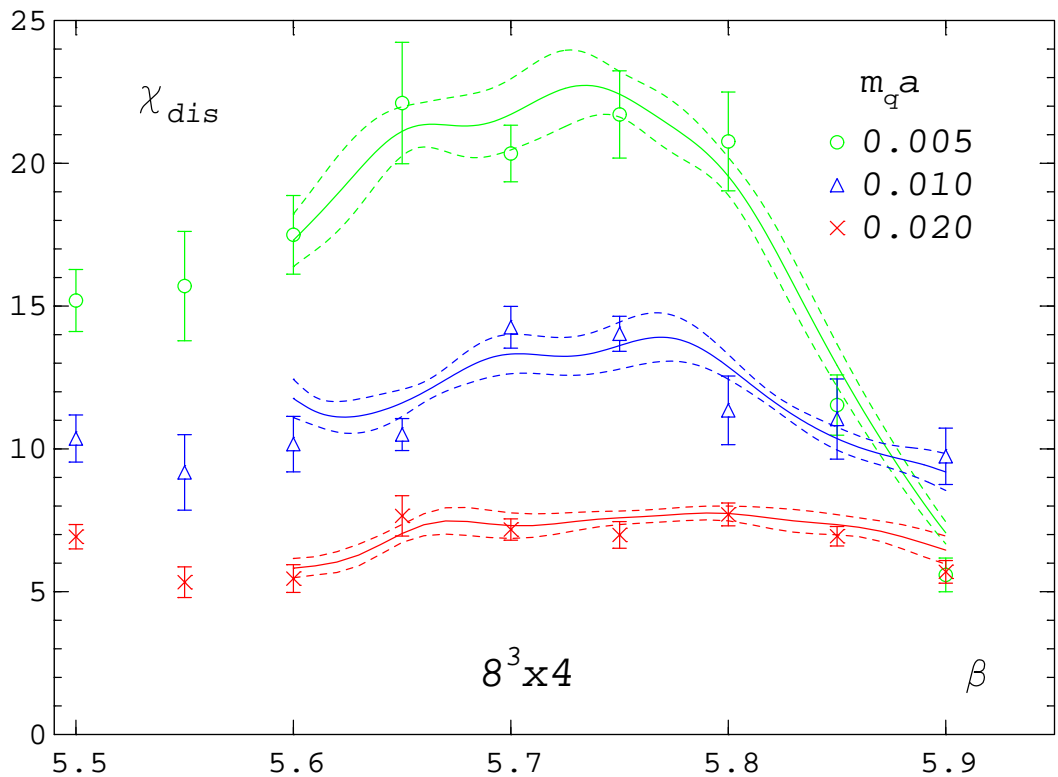, width=95mm}
\end{center}
\vspace*{0.2cm}
\caption{The susceptibility $\chi_{dis}$ as a function of $\beta$ for the
three smallest masses $m_qa=0.005,\,0.01$ and 0.02 on the $8^3\times 4$
lattice. The solid lines are the results from reweighting, the dashed 
lines show the Jackknife error corridor.}
\label{fig:chidis}
\end{figure}

%%%%%%%%%%%%%%%%%%%%%%%%%%%%%%%%%%%%%%%%%%%%%%%%%%%%%%%%%%%%%%%%%%%%%%%%%%%%%%%%
We start our search by following first the usual strategy to extrapolate
the pseudocritical couplings $\beta_{pc}(m_qa)$ to $m_qa=0$ with an ansatz
equivalent to Eq.\ (\ref{pseudo}) 
\be
\beta_{pc} = \beta_c +c (m_qa)^{1/\Delta}~.
\label{pseudob}
\ee
The $\beta_{pc}$ are in principle the positions of the maxima of $\chi_m$.
In Ref.\ \cite{Karsch:1994hm} the two parts $\chi_{dis}$ and $\chi_{con}$
of $\chi_m$ have been studied seperately. It was found that the contribution
of the connected part to $\chi_m$ is a slowly varying function of $\beta$
in the critical region and that the ratio $\chi_{con}/\chi_{dis}$ decreases
with decreasing quark mass. It suffices therefore to take the peak positions
of $\chi_{dis}$ for the extrapolation to the critical point. The data for
the susceptibility $\chi_{dis}$ from the $8^3\times 4$ lattice are shown 
in Fig.\ \ref{fig:chidis} for the three smallest masses 0.005, 0.01 and 
0.02. They have been interpolated by reweighting where we had enough 
statistics. We observe a pronounced but broad maximum for the smallest 
mass 0.005 around $\beta\approx 5.73$, for $m_qa=0.01$ a less peaked
and smaller maximum around $\beta\approx 5.74$ and for $m_qa=0.02$ only
a long and low plateau between $\beta= 5.65$ and $\beta = 5.85\,$. In
Table \ref{tab:pcl} we have listed our estimates for the pseudocritical
couplings. For $m_qa=0.01$ and 0.02 they differ somewhat from the values 
obtained in Ref.\ \cite{Karsch:1998qj} because we have more data points,
the value at $m_qa=0.005$ is new. Certainly, the errors on
$\beta_{pc}(m_qa)$ are too large to lead to a precise value of the
%------------------------------------------------------------------------
\begin{table}[b!]
\begin{center}
\begin{tabular}{|c||c|c|c|}
\hline
 $m_qa$ & 0.005 & 0.010 & 0.020 \\
\hline 
$\beta_{pc}$ & 5.73(8) & 5.74(6) & 5.75(10)\\
\hline
\end{tabular}
\end{center}
\caption{The pseudocritical couplings $\beta_{pc}$ from the $8^3\times 4$
lattice.}
\label{tab:pcl}
\end{table}
%------------------------------------------------------------------------
\newpage
\n critical point $\beta_c$. On the other hand it is clear that $\beta_c$
must be definitely smaller than the $\beta_{pc}(m_qa)$ because the
susceptibility peaks only in the high temperature phase $\beta_r > 0$. If
one nevertheless tries an extrapolation with Eq.\ (\ref{pseudo}) using the 
numbers from Table \ref{tab:pcl} and $O(2)$ and $O(6)$ values for $\Delta$
one obtains $\beta_c = 5.72(16)$ and 5.71(19), respectively. 
%------------------------------------------------------------------------
%------------------------------------------------------------------------

A second possibility to locate the critical point is based on the expected 
mass dependence of the order parameter at $\beta=\beta_c$
\be
\PBP = d_c (m_qa)^{1/\delta}~.
\label{opbc}
\ee  
The method consists in a comparison of the mass dependence of the data
at each $\beta$-value with this functional form. In practice one makes
straight line fits at fixed $\beta$ of the logarithm of the order parameter
as a function of ln($m_qa$)  
\be
\ln \PBP = \ln d_c +{1 \over \delta} \ln (m_qa)~.
\label{logpsi}
\ee
In the vicinity of $\beta_c$ one should find the lowest $\chi^2/d.o.f.$
and/or the largest fit range in $m_qa$. In addition, one obtains a first
result for $1/\delta$. A method of this type was successfully applied to
the corresponding finite size scaling problem in $SU(2)$ gauge theory 
\cite{Engels:1995em}. In Table \ref{tab:bc} we present the results for 
all fits in the interval $5.5 \le \beta \le 5.8$. At each $\beta$-value we 
have nine data points in the range $0.005 \le m_qa \le 0.10$. Because we
have two fit parameters we used at least the data points from the four 
smallest masses, if possible more, up to all the nine points.  
%------------------------------------------------------------------------
\begin{table}[t]
  \begin{center}
    \begin{tabular}{|c||c|c|c|c|c||c|}
      \hline 
 $\beta$ & $\ln d_c$ & $1/\delta$ & $m_qa$-range & $N_p$ &
 $\chi^2_{min}/d.o.f.$ &  $\chi^2_{max}/d.o.f.$ \\
      \hline
 5.80  & 1.510(37) & 0.4246(97) & 0.005 - 0.03 & 4 & 7.51 & 32.63 \\ \hline
 5.75  & 1.336(25) & 0.3549(65) & 0.005 - 0.03 & 4 & 4.44 & 28.00 \\ \hline
 5.70  & 1.124(23) & 0.2786(60) & 0.005 - 0.03 & 4 & 1.05 & 10.43 \\ \hline 
 5.65  & 1.026(09) & 0.2322(26) & 0.005 - 0.06 & 7 & 1.64 &  3.00 \\ \hline 
 5.60  & 0.949(04) & 0.1921(17) & 0.005 - 0.10 & 9 & 5.25 &  9.11 \\ \hline 
 5.55  & 0.822(19) & 0.1423(50) & 0.005 - 0.04 & 5 & 4.18 &  7.51 \\ \hline 
 5.50  & 0.755(19) & 0.1083(46) & 0.005 - 0.03 & 4 & 5.62 &  8.79 \\ \hline 
    \end{tabular}
    \caption{The parameters of the fits according to Eq.\ (\ref{logpsi}).
    The number $N_p$ of fitted data points and the fit range correspond to
    the minimal $\chi^2/d.o.f.$ with $N_p\ge 4$. The last column is the
    maximal $\chi^2/d.o.f.\,$, if $N_p$ is varied between 4 and 9.}
    \label{tab:bc}
  \end{center}
\end{table}
%------------------------------------------------------------------------
The best $\chi^2/d.o.f.$ is obtained at $\beta=5.7$ and 5.65, but for
$\beta=5.7$ only in the smallest mass interval. At $\beta=5.65$ there is 
only a small variation in $\chi^2/d.o.f.$ when $N_p$ is increased. For
$\beta=5.60$ we find a larger $\chi^2/d.o.f.$, which is due to the lowest
mass point, but the best fit includes all nine data points. The data points
at $m_qa=0.005, 0.01$ and 0.02 from the $8^3\times 4$ lattice agree here
within the errorbars with those from the $16^3\times 4$ lattice, so that 
a finite size effect is improbable. In conclusion, the range $5.6\le \beta
\le 5.7$ contains presumably the critical point $\beta_c\,$. If we look at
the corresponding $1/\delta$-values from the fits we find $0.1921\le 
1/\delta \le 0.2786$. That interval includes the value 0.2092, which is
expected from the $O(2)$ and $O(6)$ spin models, suggesting $5.6\le
\beta_c \le 5.65$.
 
\begin{figure}[t]
\begin{center}
   \epsfig{bbllx=127,bblly=265,bburx=451,bbury=588,
       file=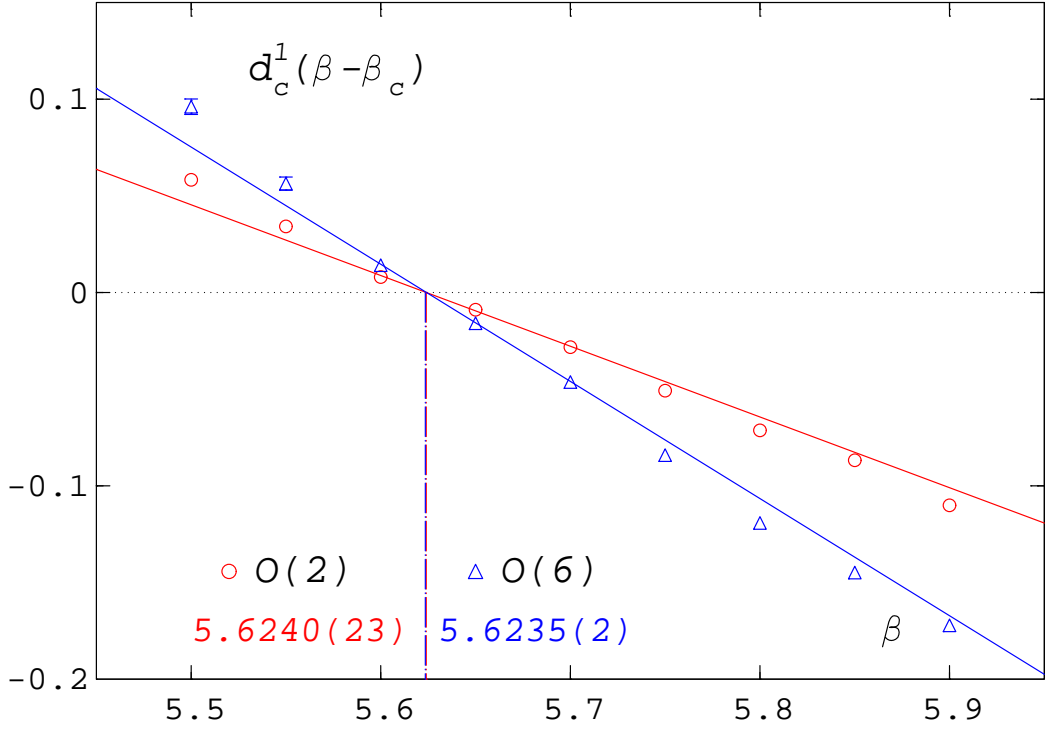,width=90mm}
\end{center}
\vspace*{0.2cm}
\caption{The parameter $d_c^1(\beta-\beta_c)$ as a function of $\beta$ for
$O(2)$ (circles) and $O(6)$ exponents (triangles). The straight lines are
fits to the values at $\beta= 5.6,5.65$ and $5.7\,$. The zeros of the fits
are given by the numbers in the plot.}
\label{fig:fitbcn}
\end{figure}

We can further improve the search for the critical point by assuming a
scaling ansatz for the order parameter 
\be
\PBP = m^{1/\delta} f(z)
\label{sca}
\ee 
as in spin models. Expanding $f(z)$ for small $|z|$ one obtains
\ba
\PBP\!\!&\! =\!&\!\!  m^{1/\delta} \left( f(0) +z f\p (0) +\dots\right) 
\label{expa} \\[0.2cm]
    \!\!&\! =\!&\!\! (m_qa)^{1/\delta} \left( d_c +d_c^1(\beta-\beta_c) 
(m_qa)^{-1/\Delta} +\dots\right)~.
\label{expan}
\ea
Here we have used the normalization condition $f(0)=1$ and Eqs.\ 
(\ref{redth}) and (\ref{opbc}). They imply $d_c = (m_qa)_0^{-1/\delta}$
and a similar relation for the amplitude $d_c^1\,$. The advantage of Eq.\
(\ref{expan}) is, that it is valid also in the vicinity of $\beta_c$ and 
at medium size values of $m_qa$. The mass should not be too small to keep
$|z|$ small and not too large to be inside the critical region where 
(\ref{sca}) is applicable. We have made fits with Eq.\ (\ref{expan}) and 
$O(2)$ and $O(6)$ exponents alternatively. The results are given in Tables
\ref{tab:o2} and \ref{tab:o6}. The shown fit range in $m_qa$ extends always
to $m_qa=0.10$. The lower limit of the range was chosen such that the
next smaller mass would increase $\chi^2/d.o.f.$ considerably. We observe
that the ranges extend to the smallest masses for $5.6\le \beta\le 5.7$.
%------------------------------------------------------------------------
\begin{table}[t]
  \begin{center}
    \begin{tabular}{|c||c|c|c|c|}
      \hline 
$\beta$ & $d_c$ & $d_c^1(\beta-\beta_c)$ & $m_qa$-range &
 $\chi^2/d.o.f.$ \\      \hline
 5.90  & 2.736 (8) & -0.1101 (16) & 0.03 - 0.10 & 2.19 \\ \hline
 5.85  & 2.704 (8) & -0.0868 (14) & 0.02 - 0.10 & 1.53 \\ \hline
 5.80  & 2.706 (6) & -0.0713 (12) & 0.02 - 0.10 & 1.16 \\ \hline
 5.75  & 2.689 (6) & -0.0508 (11) & 0.02 - 0.10 & 1.30 \\ \hline
 5.70  & 2.662 (6) & -0.0283 (10) & 0.01 - 0.10 & 1.56 \\ \hline 
 5.65  & 2.654 (5) & -0.0089 (08) & 0.005 - 0.10 & 1.68 \\ \hline 
 5.60  & 2.659 (7) & ~0.0080 (13) & 0.01 - 0.10 & 2.97 \\ \hline 
 5.55  & 2.624 (9) & ~0.0342 (20) & 0.03 - 0.10 & 2.30 \\ \hline 
 5.50  & 2.606(13) & ~0.0583 (24) & 0.03 - 0.10 & 1.99 \\ \hline 
    \end{tabular}
    \caption{The parameters for the fit (\ref{expan}) with
      $O(2)$ exponents.}
    \label{tab:o2}
\vspace*{0.4truecm}
    \begin{tabular}{|c||c|c|c|c|}
      \hline 
$\beta$ & $d_c$ & $d_c^1(\beta-\beta_c)$ & $m_qa$-range & 
$\chi^2/d.o.f.$ \\      \hline
 5.90 & 2.862 (8) & -0.1722 (16) & 0.02 - 0.10 & 2.43 \\ \hline
 5.85 & 2.811 (9) & -0.1449 (21) & 0.01 - 0.10 & 2.89 \\ \hline
 5.80 & 2.793 (6) & -0.1192 (17) & 0.01 - 0.10 & 0.75 \\ \hline
 5.75 & 2.748 (6) & -0.0841 (15) & 0.01 - 0.10 & 1.20 \\ \hline
 5.70 & 2.693 (6) & -0.0463 (14) & 0.005 - 0.10 & 2.39 \\ \hline 
 5.65 & 2.669 (6) & -0.0159 (15) & 0.005 - 0.10 & 1.73 \\ \hline 
 5.60 & 2.647 (8) & ~0.0141 (22) & 0.01 - 0.10 & 2.70 \\ \hline 
 5.55 & 2.585(11) & ~0.0563 (34) & 0.03 - 0.10 & 2.07 \\ \hline 
 5.50 & 2.538(15) & ~0.0961 (40) & 0.03 - 0.10 & 1.48 \\ \hline 
    \end{tabular}
    \caption{The parameters for the fit (\ref{expan}) with
      $O(6)$ exponents.}
    \label{tab:o6}
  \end{center}
\end{table}
%------------------------------------------------------------------------
The result for the amplitude $d_c$ should of course be independent of the
used $\beta$-value as long as $\beta$ is in the critical region. For $O(2)$
exponents (Table \ref{tab:o2}) the amplitude result is generally less
varying than for $O(6)$ exponents (Table \ref{tab:o6}), in the range 
$5.6\le \beta\le 5.7$ the result for $O(2)$ is actually constant within
error bars. If we consider this $\beta$-range as the effective critical 
region we obtain the value $d_c=2.66(1)$ for $O(2)$ and $d_c=2.67(3)$ for
$O(6)$ exponents. The assessment of the range $5.6\le \beta\le 5.7$ as the 
critical region is further supported by the parameter $d_c^1(\beta-\beta_c)$,
because it has a zero in this region. In Fig.\ \ref{fig:fitbcn} we have
plotted the results for this parameter and straight line fits to the 
three values in the critical region. We obtain coinciding values for 
$\beta_c$: 5.6240(23) for the $O(2)$ and 5.6235(2) for the $O(6)$ case.
The errors on these results should however not be overestimated, because
they are only those from the straight line fits. Nonetheless we have 
arrived at a remarkably clear estimate for $\beta_c\,$.    
%------------------------------------------------------------------------
%------------------------------------------------------------------------
%%%%%%%%%%%%%%%%%%%%%%%%%%%%%%%%%%%%%%%%%%%%%%%%%%%%%%%%%%%%%%%%%%%%%%%%%%%%%%%%
From $\beta_c$ and $\beta_d$ we can derive the ratio $T_c/T_d$
of the critical temperatures approximately with the 2-loop $\beta$-function 
\be
a\Lambda_L \approx R(\beta) = \left({ \beta \over 6b_0} \right)^{b_1/2b_0^2}
\exp \left(- {\beta \over 12 b_0} \right)~.
\label{bfunc}
\ee
The coefficients $b_0$ and $b_1$ for aQCD are given by \cite{Caswell:1974gg}
\be
b_0 = {3 \over 16 \pi^2}~, \quad b_1 = -  {90 \over (16\pi^2)^2}~,
\label{coeff}
\ee
and the temperature ratio becomes
\be
{T_c \over T_d} = {a(\beta_d) \over a(\beta_c)} \approx \left( {\beta_c \over
\beta_d} \right)^5 \exp \left ( {4 \pi^2 \over 9} (\beta_c-\beta_d)\right)
=7.8(2)~.
\label{tratio}
\ee
Though we have a lower estimate for $\beta_c$ than Ref.\ \cite{Karsch:1998qj}
we nevertheless arrive at about the same ratio. That is due to the different
value for $b_1$, which was wrongly taken as +$115/(384\pi^4)$ in 
\cite{Karsch:1998qj}.
%\clearpage

%%%%%%%%%%%%%%%%%%%%%%%%%%%%%%%%%%%%%%%%%%%%%%%%%%%%%%%%%%%%%%%%%%%%%%%%%%%%%%%%

\subsection{The Goldstone effect}
\label{sub:puregold}

\begin{figure}[b]
\begin{center}
   \epsfig{bbllx=127,bblly=265,bburx=451,bbury=588,
       file=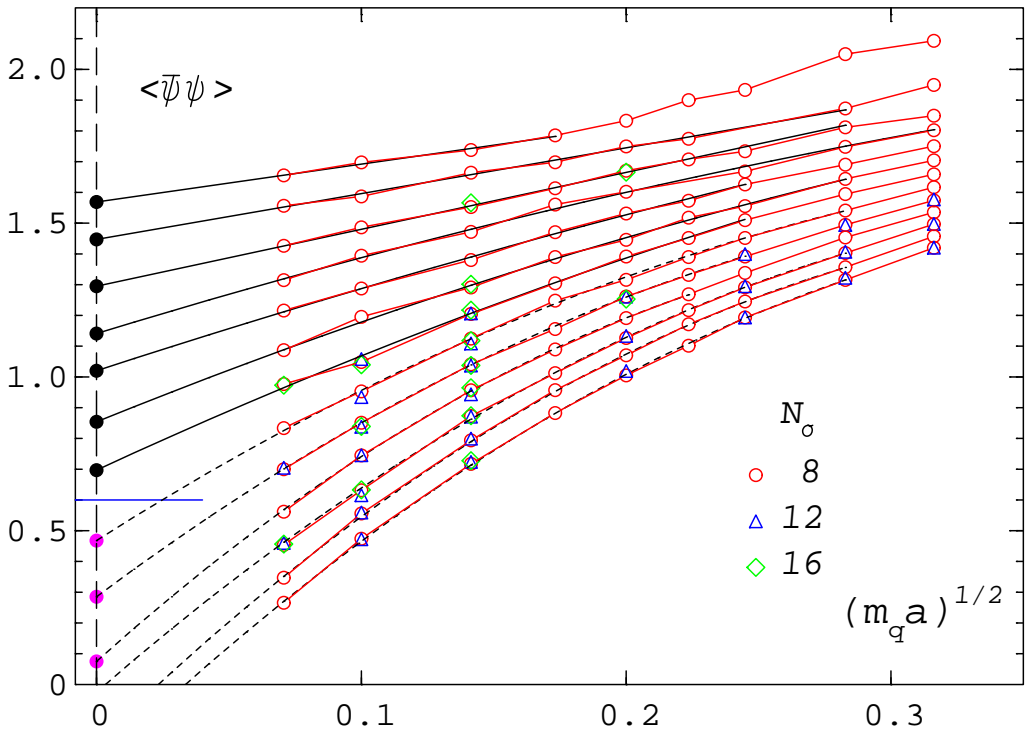,width=90mm}
\end{center}
\vspace*{0.2cm}
\caption{The chiral condensate $\PBP$ as a function of $(m_qa)^{1/2}$ for
all $\beta$-values between 5.3 (highest values) and 5.9 (lowest values)
from  $\ns^3 \times 4$ lattices with $\ns=8$ (circles), 12 (triangles) and
16 (diamonds). The lines are fits with the ansatz (\ref{pbpgold}), the
filled circles denote the extrapolations to $m_qa =0$.}
\label{fig:golda}
\end{figure}

%%%%%%%%%%%%%%%%%%%%%%%%%%%%%%%%%%%%%%%%%%%%%%%%%%%%%%%%%%%%%%%%%%%%%%%%%%%%%%%%
\n We have just successfully applied the scaling ansatz (\ref{sca}) to the 
chiral condensate data. One might therefore directly try to compare the
data with the universal scaling functions $f_G(z)$ of the spin models.
Before we can do so, however, we have to fix the normalization $\beta_0$ of
the reduced temperature $\beta_r$ (the normalization of $m$ is already known
from $d_c$). The amplitude $d_c^1$ does not fix $\beta_0$
because the expression $z df/dz$ in Eq.\ (\ref{expa}) is independent of
any normalization of $z$ or $\beta_r$. For that purpose the critical
amplitude of the chiral condensate on the coexistence line has to be 
determined. In the case of the three-dimensional $O(N)$ spin models the 
magnetization data for $T<T_c$ and $H>0$ have been extrapolated to $H=0$
using Eq.\ (\ref{mdiv}). We can proceed in the same manner to obtain
$\PBP$-values at $m_qa=0$, supposed the corresponding Goldstone ansatz 
\be
\PBP(\beta,m_qa) = \PBP(\beta,0) +c_1(\beta) (m_qa)^{1/2} +c_2(\beta) (m_qa)
+ \dots
\label{pbpgold}
\ee
for $\beta< \beta_c$ is meaningful at all. In Fig.\ \ref{fig:golda} we 
have plotted the chiral condensate data for all $\beta$-values between
5.3 and 5.9 versus $(m_qa)^{1/2}$. Here, also the available data
%------------------------------------------------------------------------
%------------------------------------------------------------------------
\n  from the $12^3\times 4$ and $16^3\times 4$ lattices are shown. Since
there are no perceptible finite size effects, the correlation length must
still be small even at the lowest masses. Because we have no data for
masses below $m_qa=0.005$ we cannot see directly from Fig.\ \ref{fig:golda}
at which $\beta$ the Goldstone behaviour, Eq.\ (\ref{pbpgold}), changes to 
Eq.\ (\ref{opbc}), the behaviour at the critical point. Nevertheless
we have made fits to the data at each $\beta$-value according to Eq.\
(\ref{pbpgold}), including the first three terms. In Table
\ref{tab:goldfit} we present the results of the fits, the corresponding
curves are shown in Fig.\ \ref{fig:golda}.
%------------------------------------------------------------------------
\begin{table}[t]
  \begin{center}
    \begin{tabular}{|c||r|c|c|c|c|}
      \hline 
 $\beta$ & $\PBP(\beta,0)$ & $c_1$ & $c_2$ & $m_qa$-range 
& $\chi^2_{min}/d.o.f.$\\
      \hline
 5.30 &  1.569 (13) & 1.23 (12) & - &          0.005 - 0.03 & 0.24 \\ \hline
 5.35 &  1.447 (07) & 1.49 (04) & - &          0.005 - 0.08 & 1.33 \\ \hline
 5.40 &  1.295 (08) & 1.85 (04) & - &          0.005 - 0.08 & 0.63 \\ \hline
 5.45 &  1.141 (11) & 2.64 (11) &~-1.72 (25) & 0.005 - 0.10 & 3.03 \\ \hline 
 5.50 &  1.020 (21) & 2.81 (26) &~-1.37 (74) & 0.005 - 0.06 & 1.19 \\ \hline
 5.55 &  0.854 (22) & 3.49 (22) &~-2.48 (54) & 0.005 - 0.08 & 2.18 \\ \hline
 5.60 &  0.697 (24) & 3.98 (29) &~-2.66 (84) & 0.005 - 0.06 & 1.67 \\ \hline
 5.65 &  0.468 (17) & 5.48 (20) &~-5.97 (54) & 0.005 - 0.08 & 1.33 \\ \hline
 5.70 &  0.285 (19) & 6.44 (22) &~-7.83 (61) & 0.005 - 0.06 & 1.28 \\ \hline
 5.75 &  0.075 (22) & 7.74 (30) & -10.77 (96) & 0.005 - 0.05 & 0.61 \\ \hline
 5.80 & -0.025 (19) & 7.51 (19) &~-8.71 (45) & 0.005 - 0.08 & 1.26 \\ \hline
 5.85 & -0.190 (12) & 8.39 (14) & -10.34 (37) & 0.005 - 0.08 & 1.14 \\ \hline
 5.90 & -0.267 (10) & 8.27 (12) &~-9.46 (31) & 0.005 - 0.08 & 1.97 \\ \hline
    \end{tabular}
    \caption{The parameters of the fits according to the Goldstone ansatz
    (\ref{pbpgold}). The upper fit boundary was fixed by the minimal
    $\chi^2/d.o.f.\,$ given in the last column.}
    \label{tab:goldfit}
  \end{center}
\end{table}
%------------------------------------------------------------------------
For $\beta=5.3$, 5.35 and 5.4 the parameter $c_2$ is zero within error bars.
At $\beta=5.3$ the influence of the nearby deconfinement transition can
explicitly be seen at the higher masses $m_qa \ge 0.04\,$, at $\beta=5.35$
for $m_qa=0.1$. The fitcurves for $\beta < \beta_c$ are shown as solid
lines. The short horizontal line in Fig.\ \ref{fig:golda} separates the
extrapolated points below and above the critical point. Obviously, the data
meet the Goldstone expectations for a three-dimensional theory quite well.

The fitcurves for $\beta > \beta_c$ are shown as dashed lines and since they
reproduce the data for $m_qa\ge 0.005$ as well with a small $\chi^2/d.o.f.$,     
the fits as such provide no clear sign for the location of $\beta_c$. 
However, they provide an upper limit for $\beta_c$ insofar as $\PBP$ at
$m_qa=0$ should be positive. From Table \ref{tab:goldfit} we see that this
is not the case for $\beta > 5.8\,$. At $\beta\approx 5.8$ the extrapolation
is zero and because of that $\beta_c$ was located near to 5.8 in Ref.\
\cite{Karsch:1998qj}. A hint to the genuine critical point is provided by
the $\beta$-dependence of the parameter $c_2$. With increasing coupling
$c_2$ increases slowly, at $\beta=5.65$ however it jumps suddenly to a
higher value indicating the change of the behaviour from Eq.\ 
(\ref{pbpgold}) to Eq.\ (\ref{opbc}). We have also tried fits for 
$5.3\le\beta<\beta_c$ with a Goldstone ansatz corresponding to a 
four-dimensional theory
\be
\PBP(\beta,m_qa) = \PBP(\beta,0) +d_1(\beta)\left[(m_qa)\ln(m_qa) -
 (m_qa)\right] + \dots~.
\label{fourdim}
\ee
The data are not well reproduced with such an ansatz and the respective 
$\chi^2/d.o.f.\,$-values are about 10-50 times larger than those in Table
\ref{tab:goldfit}. That is we confirm that aQCD at finite temperatures
in the range $T_d<T<T_c$ behaves as an effective three-dimensional 
spin model.

\begin{figure}[b]
\setlength{\unitlength}{1cm}
\begin{picture}(13,7.5)
\put(0.4,0.5){
   \epsfig{bbllx=127,bblly=264,bburx=451,bbury=587,
       file=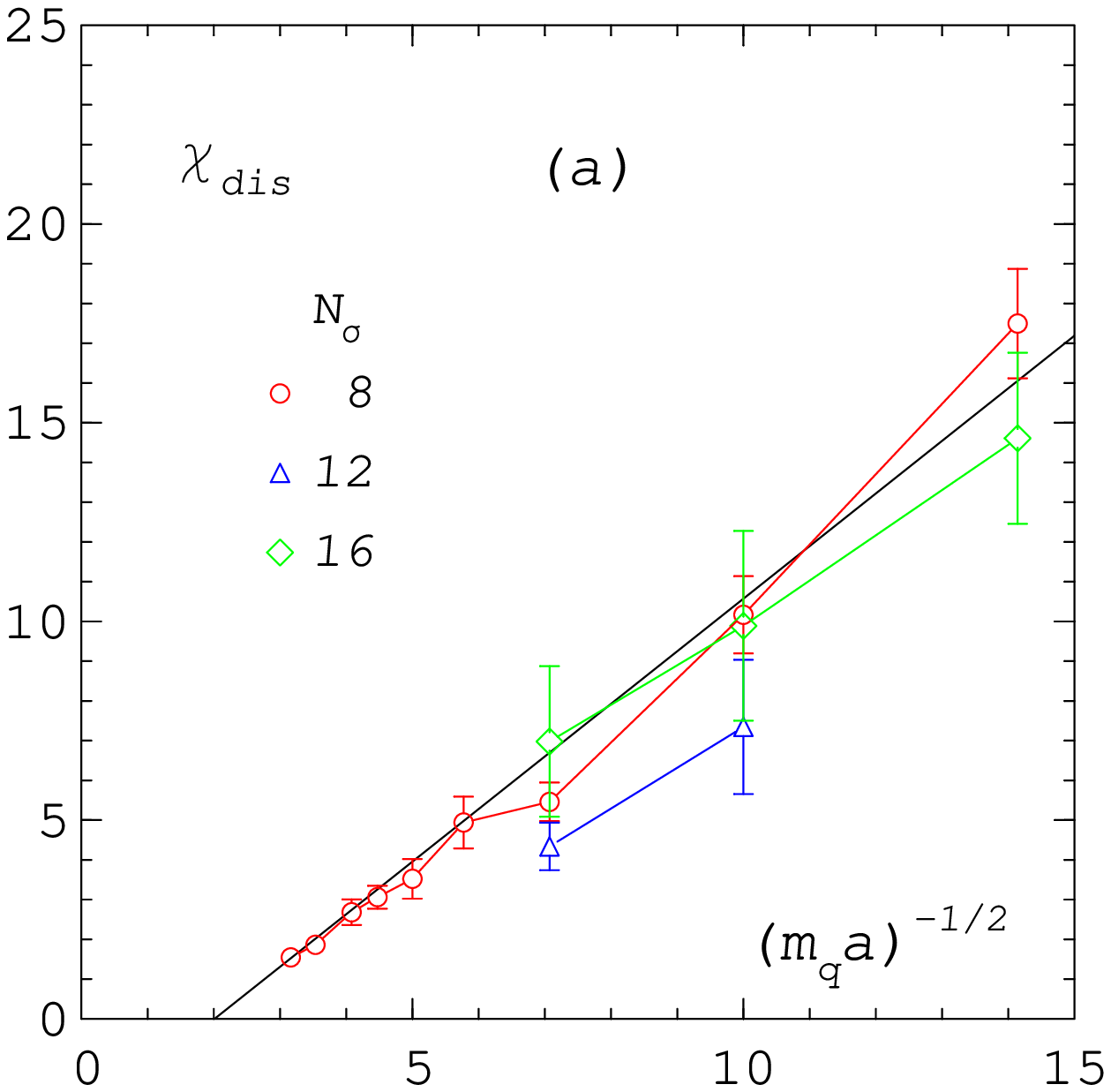, width=65mm}
          }
\put(8.0,0.5){
   \epsfig{bbllx=127,bblly=264,bburx=451,bbury=587,
       file=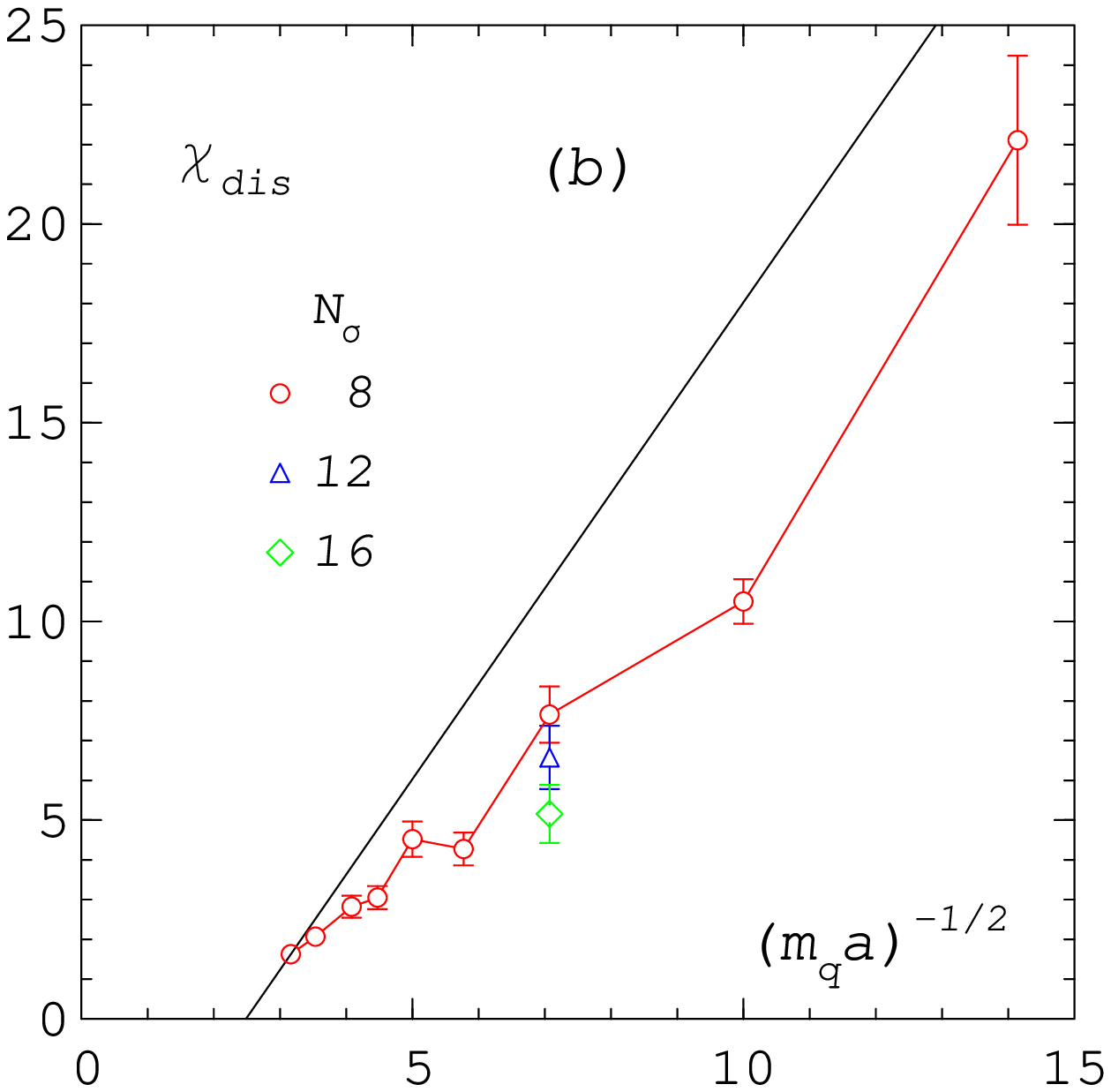, width=65mm}
          }
\end{picture}
\caption{The chiral susceptibility $\chi_{dis}$ versus $(m_qa)^{-1/2}$ for
   $\beta=5.6$ (a) and $\beta=5.65$ (b). The straight lines correspond to
   Eq.\ (\ref{goldsus}) with the parameters from Table \ref{tab:goldfit}.}   
\label{fig:gold_sus}
\end{figure}

In Fig.\ \ref{fig:gold_sus} we have plotted the disconnected part of the
chiral susceptibility $\chi_{dis}\,$ for $\beta=5.6$ and 5.65, that is 
just below and above the critical point $\beta_c\,$. A clear finite size
effect is not visible, possibly because our statistics are too low.
However, we notice a change in the mass dependence. Below the critical 
point $\chi_{dis}$ seems to follow the behaviour expected for $\chi_m$ 
from the Goldstone ansatz (\ref{pbpgold})
\be
\chi_m (\beta, m_qa) = c_2 + {c_1 \over 2 } (m_qa)^{-1/2}~, 
\label{goldsus}
\ee
whereas above $\beta_c$ the increase in the susceptibility as a function
of the mass seems to be stronger. A similar change in the mass dependence 
of $\chi_{dis}$ from $\beta=5.5$ to 5.7 can be seen in Fig.\ 4 of Ref.\ 
\cite{Karsch:1998qj}. For comparison we show in Fig.\ \ref{fig:gold_sus}
also the predictions for $\chi_m$ following from Eq.\ (\ref{goldsus}) and
the parameters from Table \ref{tab:goldfit}. At $\beta=5.6$ the agreement
with the data suggests that here $\chi_{con}$ is negligible.
%------------------------------------------------------------------------
%------------------------------------------------------------------------

\begin{figure}[t]
\setlength{\unitlength}{1cm}
\begin{picture}(13,7.5)
%\put(0,7.5){\line(1,0){15}}
\put(0.6,0.5){
   \epsfig{bbllx=127,bblly=264,bburx=451,bbury=587,
       file=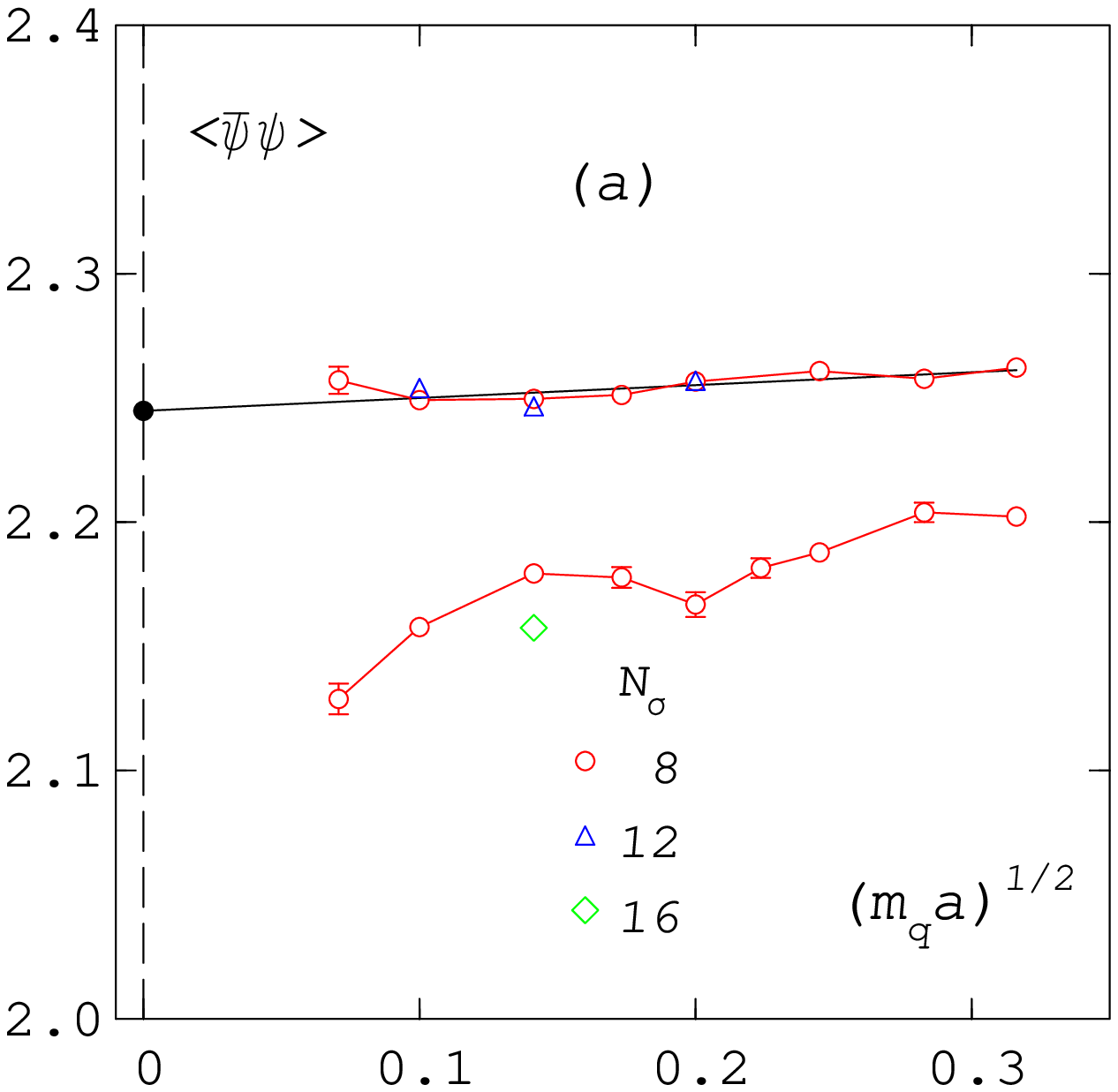, width=65mm}
          }
\put(8.2,0.5){
   \epsfig{bbllx=127,bblly=264,bburx=451,bbury=587,
       file=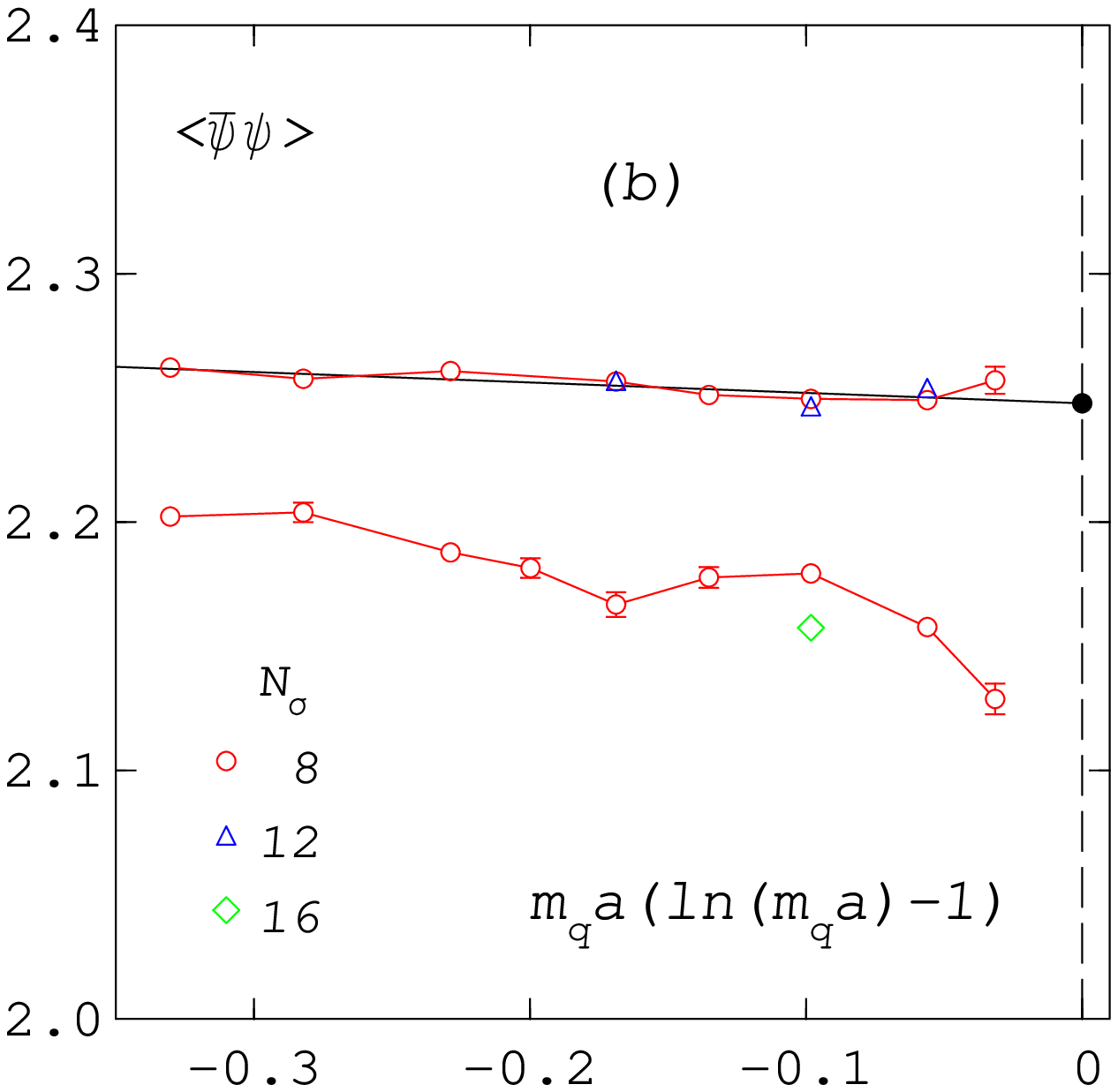, width=65mm}
          }
\end{picture}
\caption{The chiral condensate $\PBP$ for $\beta=5.1$ (upper points) 
and $\beta=5.2$ (lower points) versus $(m_qa)^{1/2}$ (a) and versus
$m_qa(\ln(m_qa) -1)$ (b). The straight lines are fits to ansatz
(\ref{pbpgold}) in (a) and to ansatz (\ref{fourdim}) in (b).}   
\label{fig:below}
\end{figure}

We have also investigated the Goldstone effect at $\beta=5.1$ and $5.2\,$.
Both values are below the deconfinement transition point $\beta_d=
5.236(3)\,$. In Fig.\ \ref{fig:below} we show the data for the chiral  
condensate. Due to the nearby deconfinement transition the data for
$\beta=5.2$ are still too noisy to enable an analysis. At $\beta=5.1$, 
on the other hand, the data are nearly independent of the mass and 
fits to ansatz (\ref{pbpgold}), Fig.\ \ref{fig:below}(a), and ansatz
(\ref{fourdim}), Fig.\ \ref{fig:below}(b), are both possible, without
favouring one of them.
%------------------------------------------------------------------------
%------------------------------------------------------------------------

\begin{figure}[t]
\begin{center}
   \epsfig{bbllx=127,bblly=265,bburx=451,bbury=588,
       file=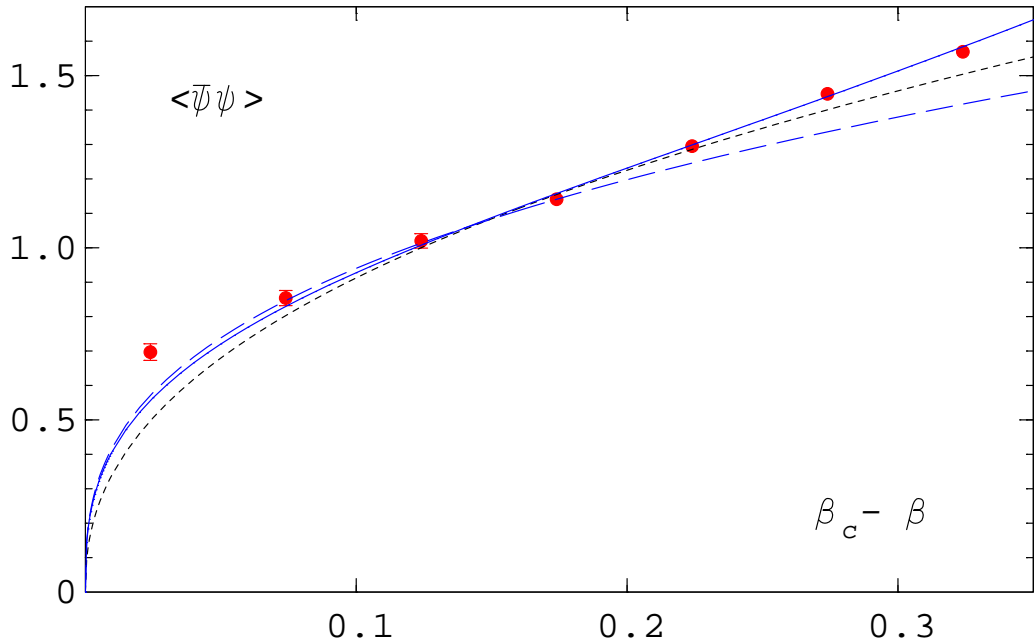,width=90mm}
\end{center}
\vspace*{-0.1cm}
\caption{The chiral condensate $\PBP$ at $m_qa=0$ from Table \ref{tab:goldfit} 
(filled circles) as a function of $\beta_c-\beta$. The dashed lines are fits
to Eq. (\ref{coex}) using the points in the range $0.05<\beta_c -\beta<0.2$
with $\beta_m$ from $O(2)$ (long dashes) and $O(6)$ (short dashes). The solid
line is a fit to Eq.\ (\ref{correc}) with $\beta_m$ fixed to the
$O(2)$-value.}
\label{fig:pbpcoex}
\end{figure}

As mentioned already, the verification of the Goldstone ansatz 
(\ref{pbpgold}) leads to an estimate of the chiral condensate on the 
coexistence line. In the vicinity of the critical point we expect the 
following critical behaviour 
\be
\PBP(\beta<\beta_c, m_qa=0) = B (\beta_c-\beta)^{\beta_m} = 
(-\beta_r)^{\beta_m}~. 
\label{coex}
\ee
Here, $\beta_m$ is the magnetic critical exponent (called $\beta$
in the case of the $O(N)$ spin models). From the second part of Eq.\ 
(\ref{coex}) it is clear that the critical amplitude $B$ and the
normalization $\beta_0$ are related by $\beta_0=B^{-1/\beta_m}$. In order 
to determine $B$ we have to evaluate the estimates of the chiral condensate
at $m_qa=0$ from Table \ref{tab:goldfit}. These data are shown in Fig.\
\ref{fig:pbpcoex}. Of course, at larger values of $\beta_c-\beta$ 
corrections to the leading
term are expected to contribute. We have therefore first tried fits
to Eq.\ (\ref{coex}) without the last three points, using $\beta_m$ from
the $O(2)$ and $O(6)$ models, or as a free parameter. In all three cases
the $\chi^2/d.o.f.$ is of the order 16-20 when the point at $\beta=5.6$
is included in the fit. Most probably, due to the vicinity of the critical
point, the extrapolation of the chiral
condensate to $m_qa=0$ at $\beta=5.6$ is not as reliable as at the higher 
$\beta$-values. Discarding that point, we obtain excellent and essentially
coinciding fits, if $\beta_m$ is treated as a free parameter and for
$\beta_m(O(2))$. The fit is not as good, if $\beta_m$ from the $O(6)$ model
is used. We have also made fits with an ansatz including a
correction-to-scaling term
\be
\PBP(\beta<\beta_c, m_qa=0) = B (\beta_c-\beta)^{\beta_m}[1+ 
b_1(\beta_c-\beta)^{\theta}]~,
\label{correc}
\ee
and all available points apart from the first one. In Fig.\ \ref{fig:pbpcoex}
we show the result for $\beta_m$ fixed to the $O(2)$-value as a solid line.
Here, the correction exponent is rather large, $\theta\approx 2$. A similar
fit with
$\beta_m(O(6))$ is again worse and leads to large errors in the parameters.
A fit to Eq.\ (\ref{correc}), where also $\beta_m$ is free, is very unstable
and leads to a negative $\beta_m$. Details of our fits are listed in Table
\ref{tab:fitco}. 
%------------------------------------------------------------------------
\begin{table}[t]
  \begin{center}
    \begin{tabular}{|c|c|c|c||c|c|}
      \hline 
 $B$ & $\beta_m$ & $b_1$ & $\theta$ & $(\beta_c-\beta)$-range 
& $\chi^2/d.o.f.$\\
      \hline
 2.10(02) &  $O(2)$  & - & - & 0.074 - 0.174 & 0.075 \\ \hline
 2.43(02) &  $O(6)$  & - & - & 0.074 - 0.174 & 4.100 \\ \hline
 2.06(12) & 0.338(30) &- & - & 0.074 - 0.174 & 0.012 \\ \hline
 2.04(06) &  $O(2)$  & 1.39(86) & 1.98(68) & 0.074 - 0.324 & 2.21 \\ \hline 
 2.42(03) &  $O(6)$  & 2.4(5.3) & 3.4(2.1) & 0.074 - 0.324 & 4.06 \\ \hline
    \end{tabular}
    \caption{The parameters of the fits according to Eqs.\ (\ref{coex})
    and (\ref{correc}).}
    \label{tab:fitco}
  \end{center}
\end{table}
%------------------------------------------------------------------------
\vspace*{-0.1cm}
%------------------------------------------------------------------------
%------------------------------------------------------------------------
\newpage
%%%%%%%%%%%%%%%%%%%%%%%%%%%%%%%%%%%%%%%%%%%%%%%%%%%%%%%%%%%%%%%%%%%%%%%%%%%%%%%%
\subsection{Scaling behaviour of the chiral condensate}
\label{sub:scalepbp}
\begin{figure}[b!]
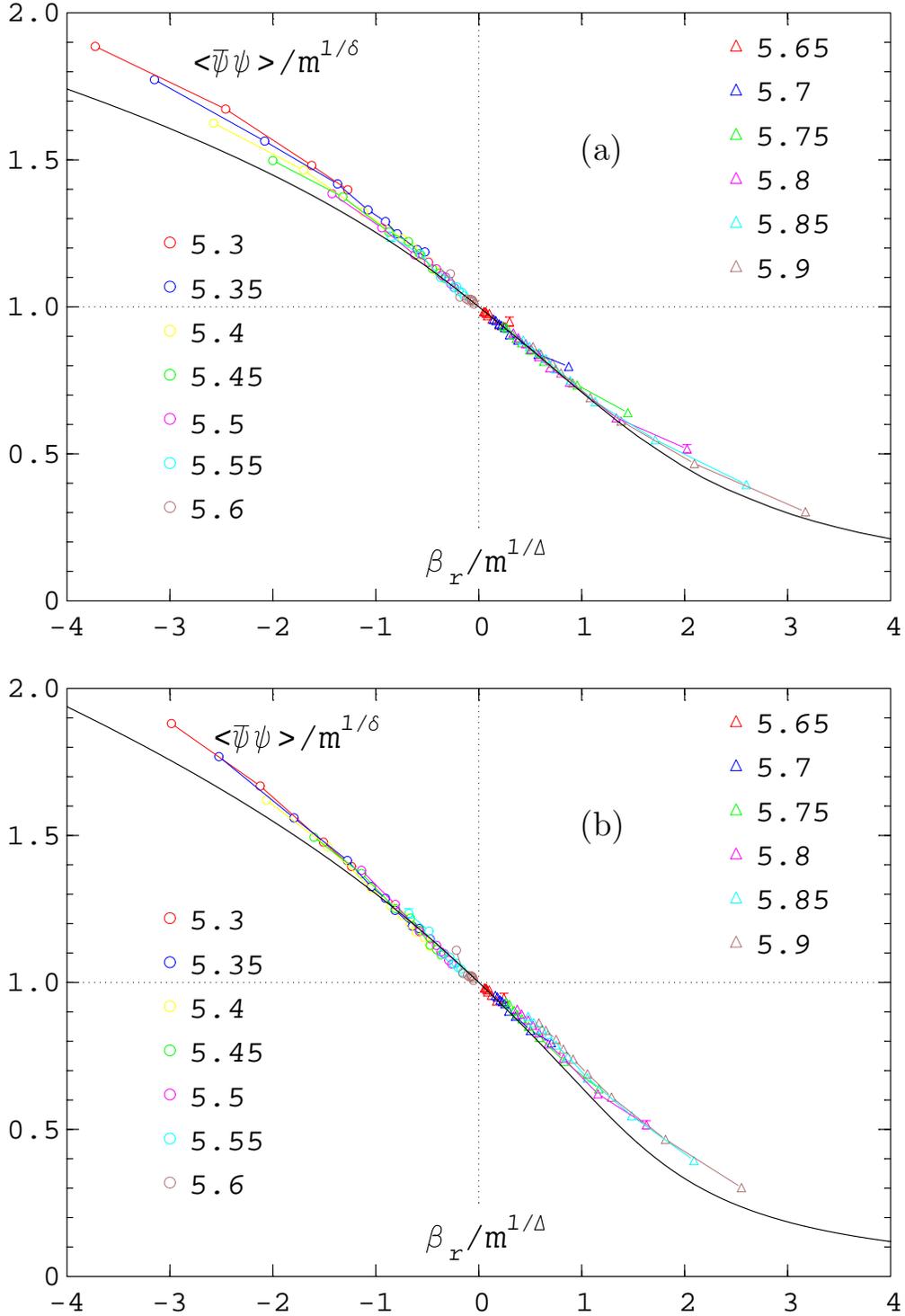

\caption{The scaling function $f(z)=\PBP/m^{1/\delta}$ versus the scaling
variable $z=\beta_r/m^{1/\Delta}$ with data from the $8^3\times 4$ lattice
for $O(2)$ (a) and $O(6)$ (b) critical exponents and normalizations from
Table \ref{tab:norm}. The solid curves are the universal scaling functions
from the respective $O(N)$ spin models. The data points are connected by
straight lines to guide the eye, the numbers indicate the $\beta$-values
of the data.}
\label{fig:scaleno}
\end{figure}

%%%%%%%%%%%%%%%%%%%%%%%%%%%%%%%%%%%%%%%%%%%%%%%%%%%%%%%%%%%%%%%%%%%%%%%%%%%%%%%%
In the last subsections we have seen that the chiral condensate data in the
close vicinity of the chiral phase transition are compatible with $O(2)$ and, 
to a lesser degree, also $O(6)$ critical behaviour. In the following we want
to compare the universal scaling function for aQCD, $f=\PBP m^{-1/\delta}$, to 
those of the spin models, $f_G=Mh^{-1/\delta}$. In Fig.\ \ref{fig:scaleno} we
show the data for $\PBP m^{-1/\delta}$ as a function of the scaling variable
$z=\beta_r/m^{1/\Delta}$ with the critical exponents for the $O(2)$ (a) and
$O(6)$ (b) models from Table \ref{tab:O(N)}. The used normalizations 
$(m_qa)_0$ and $\beta_0$ are shown in Table \ref{tab:norm}. The mass 
normalization was calculated from the amplitude $d_c$ determined in subsection 
\ref{sub:critpoint}. The normalization $\beta_0$ can be deduced from the
amplitude $B$ computed in the last subsection. Alternatively, one may fix the
derivative of the scaling function at the critical point to the one of the
corresponding $O(N)$ model. This is what we have done for the plots.
The derivative is given by 
\be
f\p(0)=d^1_c \beta_0 d_c^{~(1/\beta_m -1)}~.
\label{fprime}
\ee
The derivatives of the $O(2)$ and $O(6)$ scaling functions can be found from
the respective parametrizations in Refs.\ \cite{Engels:2000xw,Holtmann:2003he}. 
Choosing this possibility we obtain $B=2.06$ for $O(2)$ and $B=2.27$ for
$O(6)$. That is, for $O(2)$ we find a $B$-value which is in agreement with the 
$O(2)$ and free fit results from Table \ref{tab:fitco}, whereas the $O(6)$
fits lead to somewhat higher $B$-values. In Fig.\ \ref{fig:scaleno} we have
plotted only the data from the range $5.3\le \beta \le 5.9$, at $\beta=5.3$
only the data for $m_qa \le 0.03$. All other data are either distorted by the 
deconfinement transition or too far outside the critical region. 

We observe in Fig.\ \ref{fig:scaleno} (a) that the data scale very well with
$O(2)$ exponents for small $|z|$ and agree there also with the $O(2)$ scaling 
function $f_G(z)$. For decreasing $z<0$ the data for different $\beta$ start
to deviate from each other. A similar effect was found for the $O(2)$
data \cite{Engels:2000xw} and is characteristic for corrections to scaling. 
At $z>0$ the data scale well apart from the data for $m_qa=0.005$. This may 
be due to the presence of finite size effects, though the few data with low
statistics which we have from the larger lattices coincide inside the error
bars with the $8^3\times 4$ data. In Fig.\ \ref{fig:scaleno} (b), where we 
%%%%%%%%%%%%%%%%%%%%%%%%%%%%%%%%%%%%%%%%%%%%%%%%%%%%%%%%%%%%%%%%%%%%%%%%%%%%%%%%
\begin{table}[b]
  \tabcolsep1.1ex
  \begin{center}
    \begin{tabular}[b]{|c||c|c||c|c|c|c|}
      \hline
      & $(m_qa)_0$ & $\beta_0$ & $d_c$ & $d_c^1$ & $f_G\p(0)$ & $B$ \\ \hline
      $O(2)$ & 0.0093 & 0.126 & 2.66 & -0.366 & -0.286  & 2.06 \\ \hline
      $O(6)$ & 0.0091 & 0.145 & 2.67 & -0.605 & -0.332  & 2.27 \\ \hline
    \end{tabular}
    \caption{The normalization constants and the corresponding amplitudes
      used in the scaling plots.}
    \label{tab:norm}
  \end{center}
\end{table}
%%%%%%%%%%%%%%%%%%%%%%%%%%%%%%%%%%%%%%%%%%%%%%%%%%%%%%%%%%%%%%%%%%%%%%%%%%%%%%%%
show the data using $O(6)$ parameters, the situation is somewhat different.
Here, the data scale very well apart from the neighbourhood of the critical 
\newpage
%----------------------------------------------------------------------------
\n \setlength{\unitlength}{1cm}
\begin{picture}(10,18.5)
\put(3.0,10){
   \epsfig{bbllx=127,bblly=265,bburx=451,bbury=588,
       file=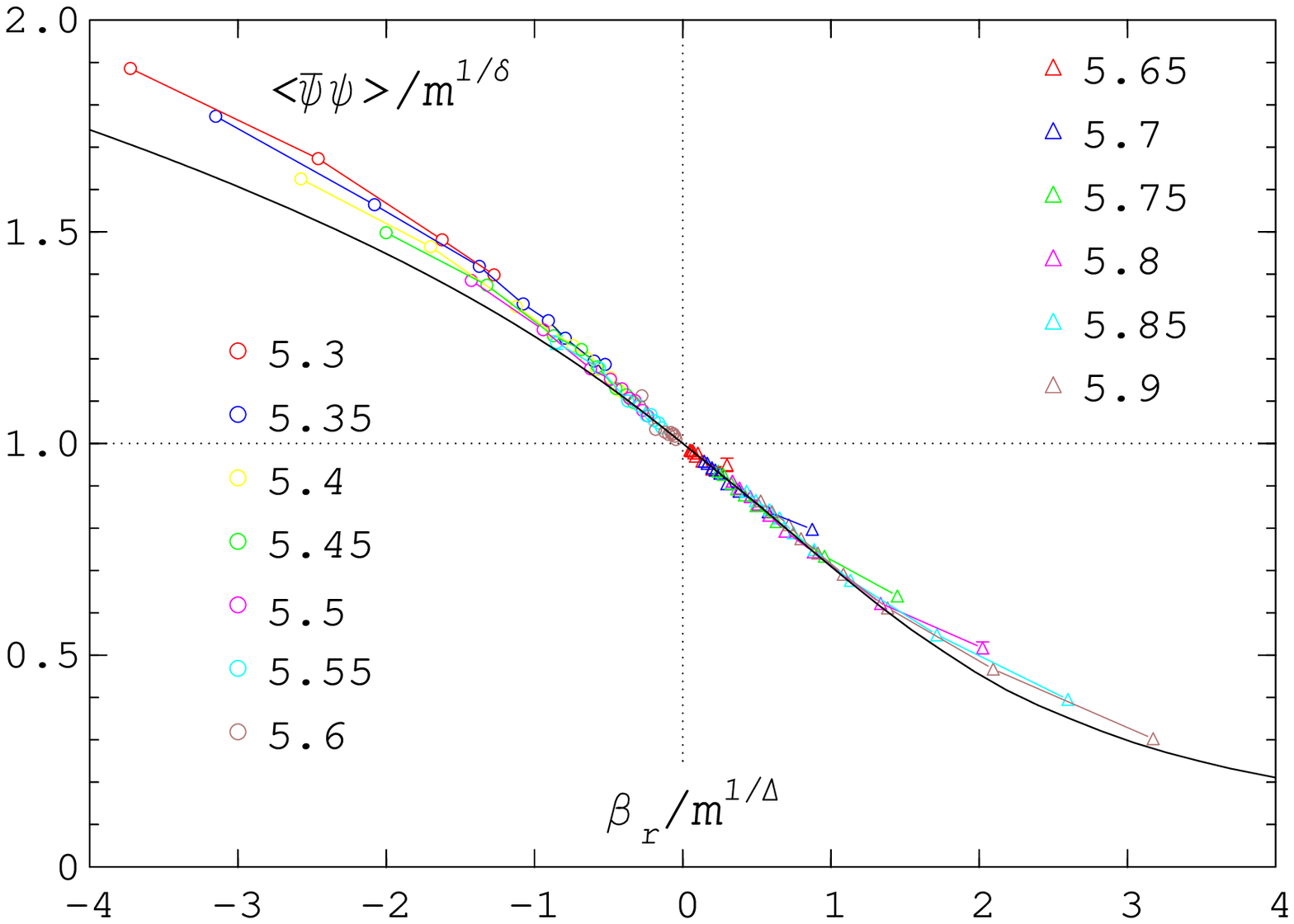, width=87mm}
          }  
\put(3.0,0){ 
   \epsfig{bbllx=127,bblly=265,bburx=451,bbury=588,
       file=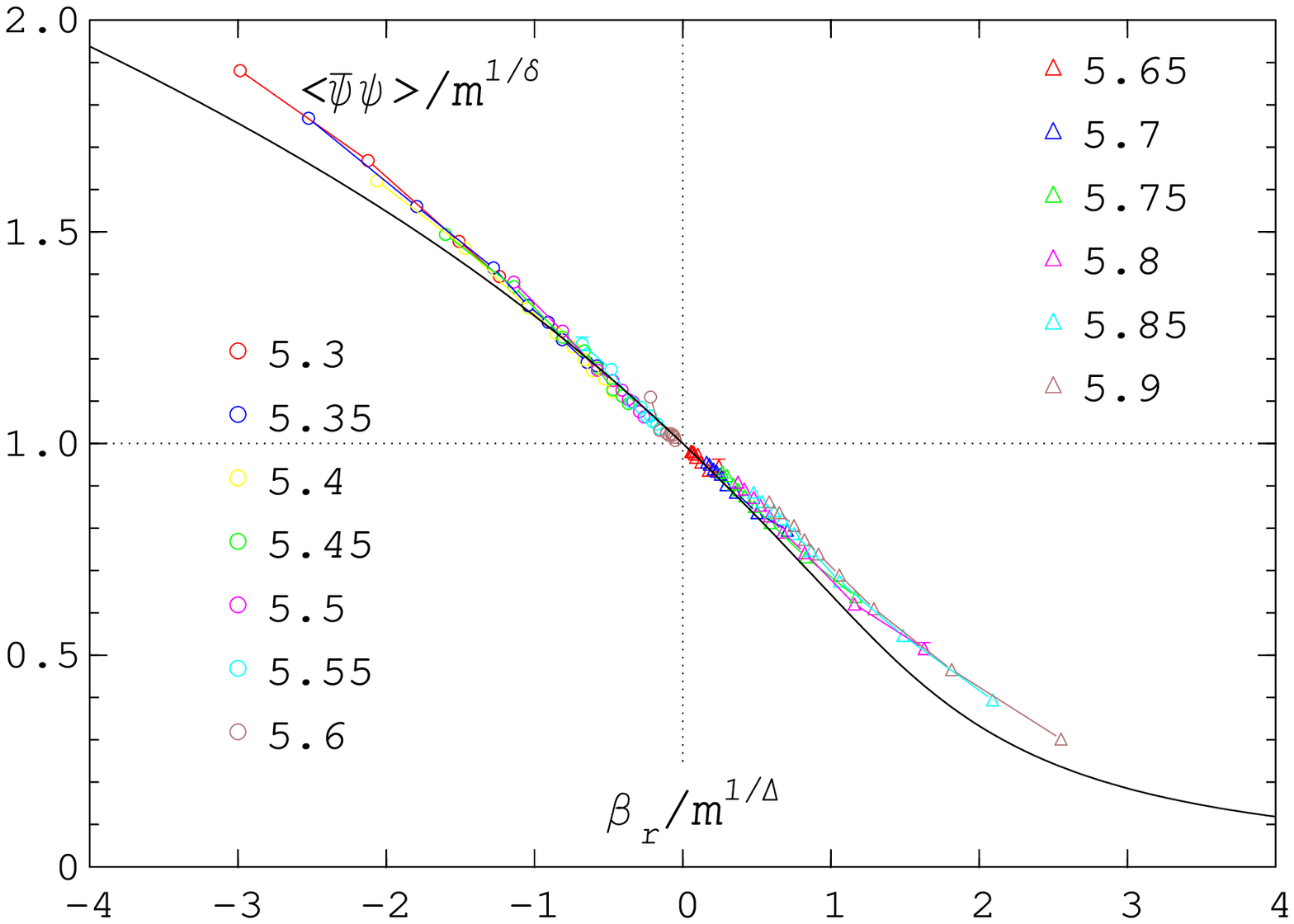, width=87mm}
          }
\put(9.0,16.5){\large{(a)}}
\put(9.0,6.5){\large{(b)}}
\end{picture}
%----------------------------------------------------------------------------
%------------------------------------------------------------------------
\newpage
\n  point where the data spread somewhat and deviate from the $O(6)$
scaling function, except very close to the normalization point $z=0$.
This is in contrast to the $O(2)$ case and can be seen in more detail in
Fig.\ \ref{fig:o2-6}, which shows the scaling behaviour in the close
vicinity of the critical point. 

%------------------------------------------------------------------------
\begin{figure}[t]
\setlength{\unitlength}{1cm}
\begin{picture}(13,7.5)
\put(0.4,0.5){
   \epsfig{bbllx=127,bblly=264,bburx=451,bbury=587,
       file=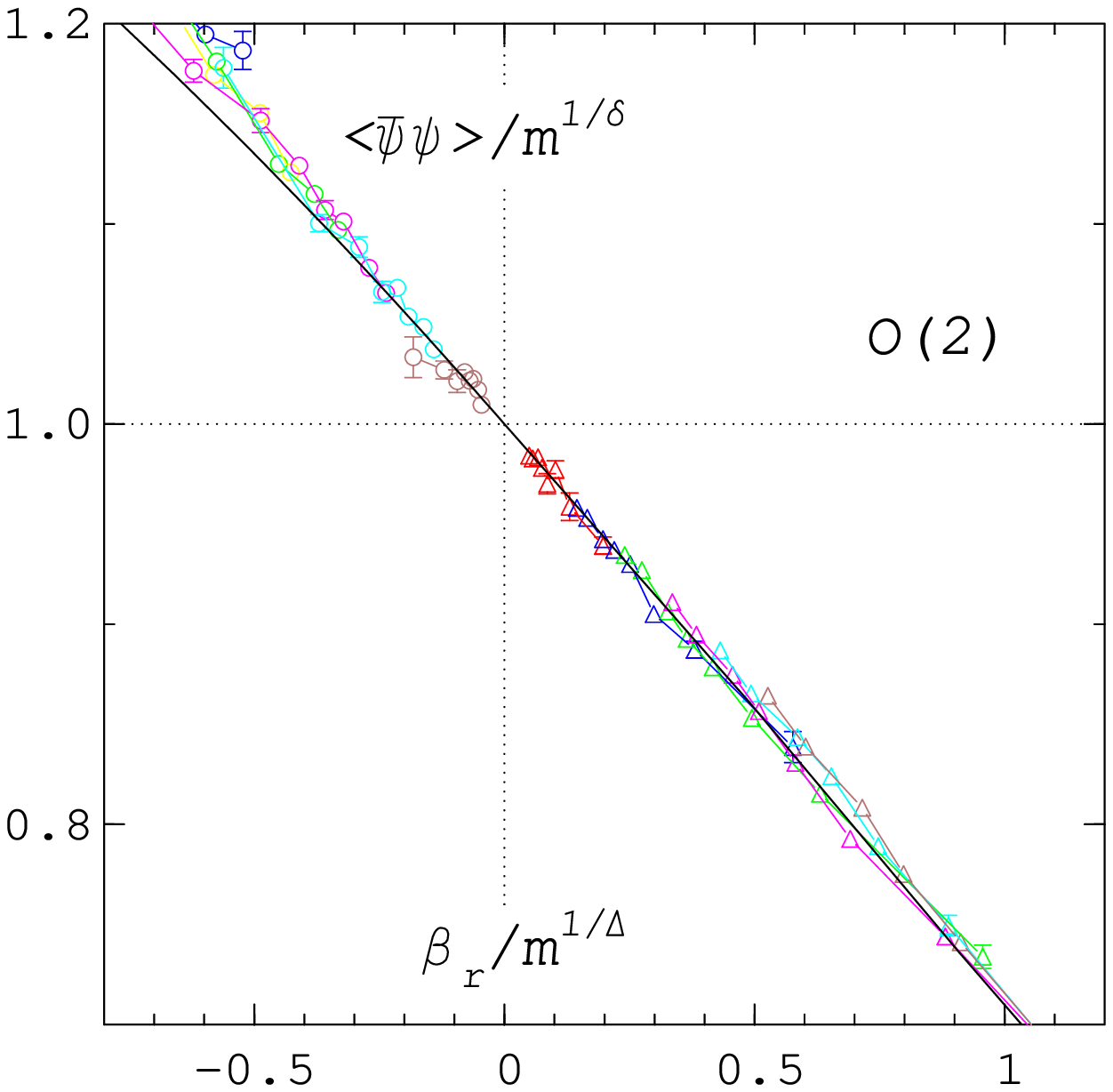, width=65mm}
          }
\put(8.0,0.5){
   \epsfig{bbllx=127,bblly=264,bburx=451,bbury=587,
       file=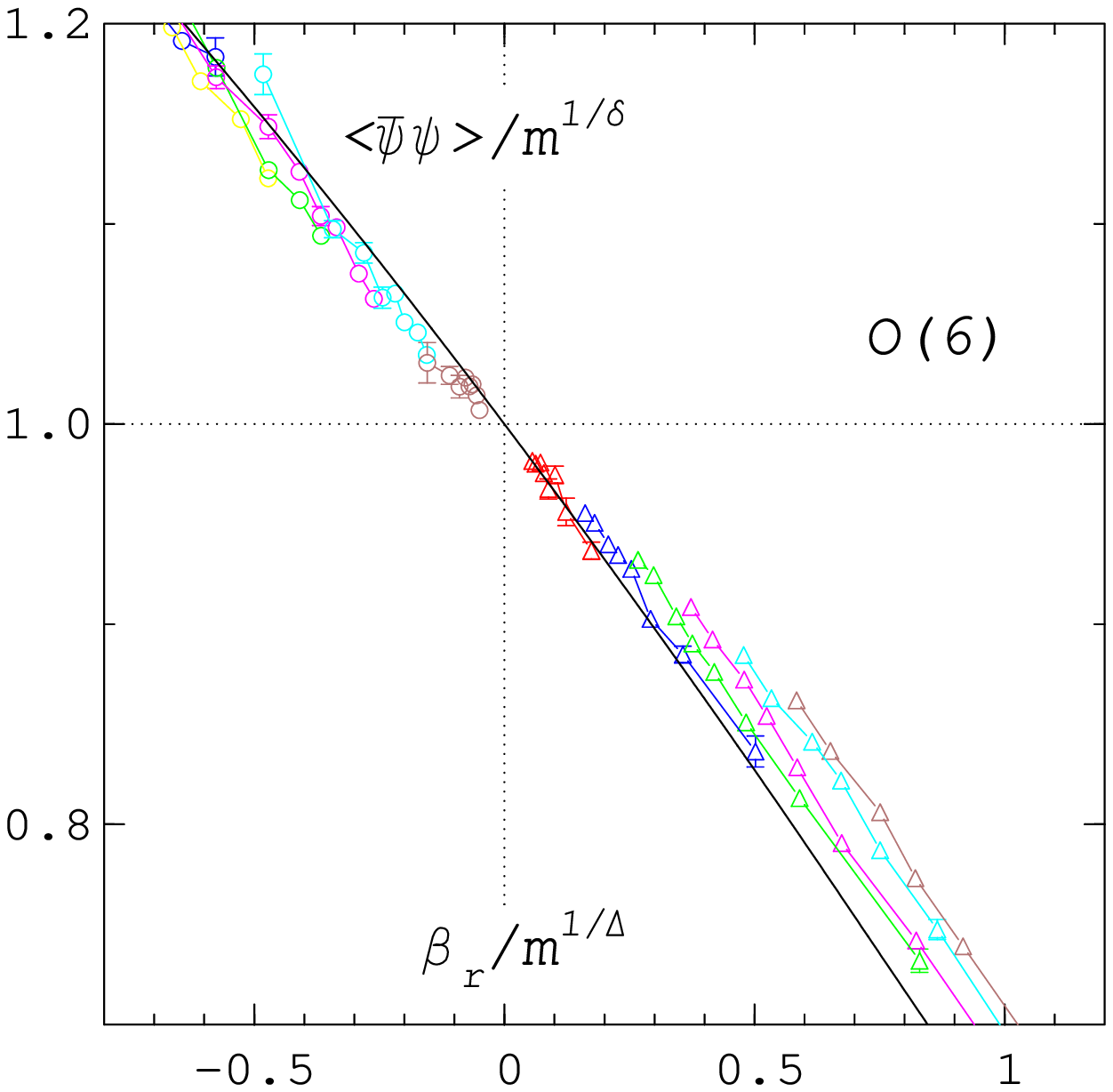, width=65mm}
          }
\end{picture}
\caption{Comparison of the scaling behaviour of $\PBP/m^{1/\delta}$
   for $O(2)$ (left) and $O(6)$ (right) exponents and normalizations in
   the vicinity of the critical point. The notation is the same as in the 
   previous figure. For clarity the data for $m_qa=0.005$ have been 
omitted. The black solid lines are the respective $O(N)$ scaling functions.}   
\label{fig:o2-6}
\end{figure}
\begin{figure}[t]
\begin{center}
   \epsfig{bbllx=127,bblly=265,bburx=451,bbury=588,
       file=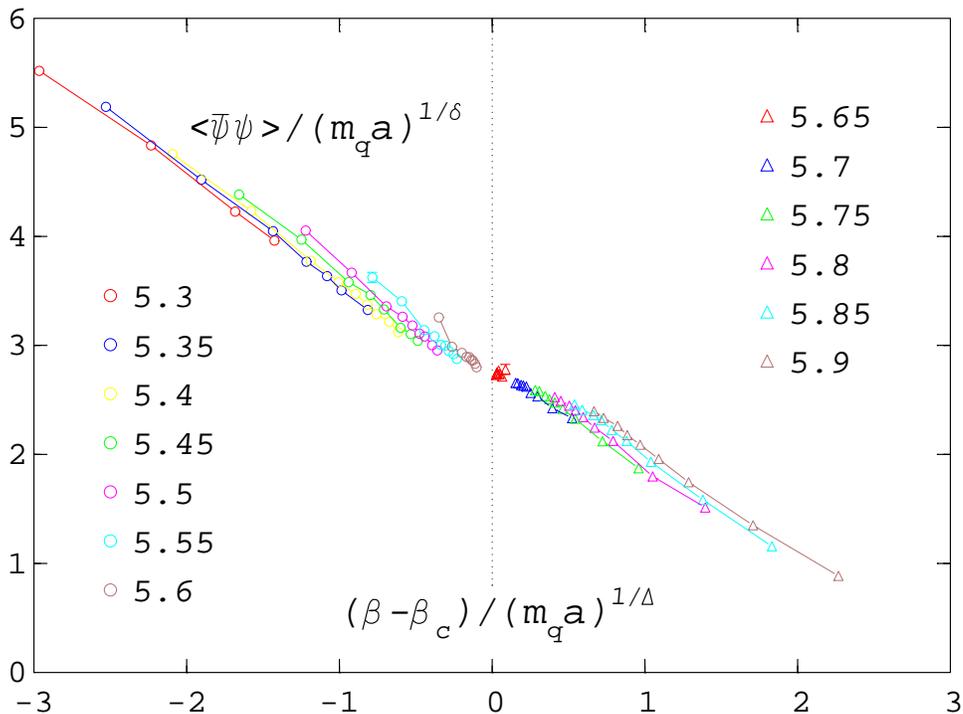, width=87mm}
\end{center}
\vspace*{0.2cm}
%----------------------------------------------------------------------------
\caption{Scaling test for $\PBP/(m_qa)^{1/\delta}$ with critical exponents
from \cite{Basile:2004wa}. The data from the $8^3\times 4$ lattice are
plotted as a function of $(\beta-\beta_c)/(m_qa)^{1/\Delta}$. 
The notation is the same as in Fig.\ \ref{fig:scaleno}.}
\label{fig:scalepel}
\end{figure}

%------------------------------------------------------------------------
As mentioned in the introduction there are first results \cite{Basile:2004wa}
for the universality class which is expected to govern the chiral transition
in the continuum limit. The universal scaling function of the class is still
unknown, but estimates of two critical exponents, namely $\eta\approx 0.2\, 
, \nu\approx 1.1$ are known. The errors are still large, e.g. $\eta=0.2(1)$
\cite{ettore}. Due to the relation 
\be
\delta={5-\eta \over 1+\eta }
\label{deleta} 
\ee 
for $d=3$
this implies $\delta=4.0(4)\,$, and from the considerations in subsection 
\ref{sub:critpoint} and the numbers in Table \ref{tab:bc} this entails a
shift in the location of the critical point. Though we have no scaling 
function to compare with we can still perform a test on the scaling of the
data assuming critical exponents compatible with the estimates of Ref.\
\cite{Basile:2004wa}. We have varied the exponents correspondingly and 
calculated each time the adequate critical point. In Fig.\ \ref{fig:scalepel}
we show the best scaling result for $\PBP/(m_qa)^{1/\delta}$ as a 
function of $(\beta-\beta_c)/(m_qa)^{1/\Delta}$ which is obtained for
$\delta=4.4\, ,\nu=1.0$ ($1/\Delta =0.409$) and $\beta_c=5.64$. We see 
from Fig.\ \ref{fig:scalepel} that the data still spread considerably
around the critical point and do not indicate a definite scaling function.
\newpage
%----------------------------------------------------------------------------
%------------------------------------------------------------------------
%%%%%%%%%%%%%%%%%%%%%%%%%%%%%%%%%%%%%%%%%%%%%%%%%%%%%%%%%%%%%%%%%%%%%%%%%%%%%%%%
\section{Summary and Conclusions}
\label{section:conclusion}
%%%%%%%%%%%%%%%%%%%%%%%%%%%%%%%%%%%%%%%%%%%%%%%%%%%%%%%%%%%%%%%%%%%%%%%%%%%%%%%%
In this paper we have studied QCD with two flavours of staggered Dirac
fer\-mions in the adjoint representation at finite temperature. Our main 
objective was the investigation of the chiral transition and the determination
of the associated universality class. For ordinary QCD with two staggered
fermions this issue is still under discussion and may be difficult to decide
because the deconfinement transition coincides with the chiral transition. In
contrast to QCD the corresponding transitions of aQCD occur at two distinct
critical temperatures and therefore allow to study the transitions separately.   
A major prerequisite of any scaling test on critical behaviour is the exact
location of the critical point. In order to achieve that goal and also to 
obtain sufficient data for the tests we have extended the existing data base
for aQCD supplied by the work of Karsch and L\"utgemeier \cite{Karsch:1998qj}.  

 The critical point of the chiral transition was then determined using several
strategies. Whereas the peak positions of the chiral susceptibility provide
only first hints on a $\beta_c$-value below 5.79 - the result of Ref.\ 
\cite{Karsch:1998qj} - the evaluation of the mass dependence of $\PBP$ leads
definitely to an interval $5.6\le \beta_c \le 5.7$ and a value $0.19\le 1/\delta
\le 0.28$ for the citical exponent of $\PBP$ on the critical line. The known
values of $\delta$ for the $O(N)$ spin models are all close 4.8, that is
$1/\delta \approx 0.21$. The value $\delta=4.0(4)$ of the new 
universality class proposed by Ref.\ \cite{Basile:2004wa} is however also
compatible with the found $1/\delta$-interval.
In the following we improved on our $\beta_c$-estimate by including the 
next term in the scaling ansatz, using the exponents of the $O(2)$ and $O(6)$
model and later also the ones from Ref.\ \cite{Basile:2004wa}. As it turned
out both spin models lead to the same rather accurate value $\beta_c=5.624(2)$.
This value and the respective value $\beta_d=5.236(3)$ \cite{Karsch:1998qj} 
of the deconfinement transition yield, via the two-loop beta function, the
ratio $T_c/T_d = 7.8(2)$.
%----------------------------------------------------------------------------
\begin{figure}[t]
\setlength{\unitlength}{1cm}
\begin{picture}(10,9.6)
\put(2.3,4.5){
   \epsfig{bbllx=127,bblly=265,bburx=451,bbury=588,
       file=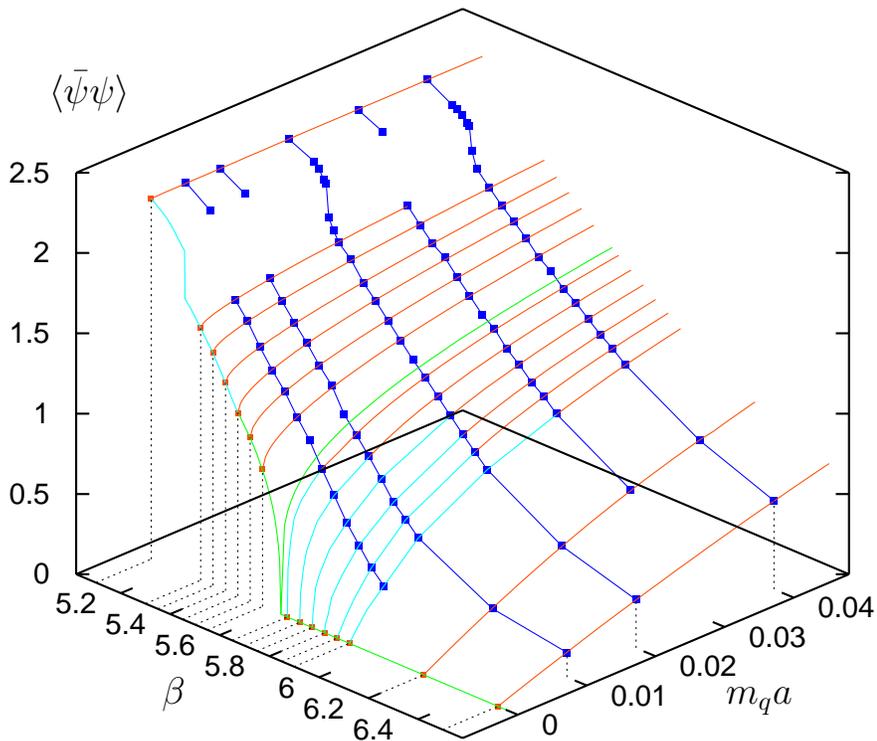, width=90mm}
          }
\put(1.8,8.5){\large$\PBP$}
\put(10.8,0.5){\large$m_qa$}
\put(3.3,0.5){\large$\beta$}
\end{picture}
\\
\caption{Three-dimensional plot of the chiral condensate $\PBP$ for all
  $\beta$-values und $m_qa\le 0.04$. The data are from the $8^3\times 4$
  lattice. The dark blue lines connect the same masses, the green ones show
  fits to the critical behaviour on the coexistence line and the critical
  line. The red lines are fits of the mass dependencies at fixed $\beta$
  and the light blue lines show the expected behaviour in regions where we
  do not have enough data. The dotted lines show the hights and indicate
  the $\beta$-values.}
\label{3d_chi_fits}
\end{figure}
%----------------------------------------------------------------------------

A further aim of our paper was the verification of the Goldstone effect
which is expected to control the mass dependence of the chiral condensate
below the chiral transition temperature. In the three-dimensional $O(N)$
models the effect is even detectable in the critical region close to $T_c$
due to its influence on the universal scaling function $f_G(z)$. We have
evaluated the chiral condensate data at fixed $\beta < \beta_c$ in the same 
manner as it was done for the magnetization of the three-dimensional $O(N)$
models and find a completely analogous behaviour, that is aQCD acts as an
effective three-dimensional $O(N)$ spin model in its chirally broken phase.   
However, close to the deconfinement transition this behaviour is disturbed
and $\PBP$ develops a gap. Below $\beta_d$, at $\beta=5.1$, $\PBP$ becomes
a smooth, though only slightly varying function of $m_qa$, which could as
well be interpreted as the behaviour of an effective four-dimensional theory.  

Using the obtained values for the critical point and the critical amplitudes
for the normalization of the scales we have carried out scaling tests and 
comparisons to the scaling functions of the $O(2)$ and $O(6)$ models. With
the $O(2)$ critical exponents the data are scaling convincingly in the direct
neighbourhood of the critical point, that is for small values of $|z|$ and
they coincide there with the $O(2)$ scaling function. For larger values of 
$(-z)$ the data show - like in the original non-linear $\sigma$-model - 
increasing deviations, which are typical for corrections to scaling. 
Using the $O(6)$ parameters for the scaling test seems at first sight to
lead to scaling of the data in a larger region of $z$. A closer inspection
in the vicinity of the critical point, where the $O(2)$ test works, shows
however, that the data spread there and deviate somewhat from the $O(6)$
scaling function. This observation is confirmed by the calculation of the
amplitude $B$ from the derivative of the scaling function: it is in better 
agreement with the value found from the Goldstone effect for $O(2)$ than
for $O(6)$. We have also carried out scaling tests with the critical
exponents calculated in Ref.\ \cite{Basile:2004wa} for the universality class
of the continuum limit. Our lattice data do not yet support the proposed
class. 

In conclusion, our findings have led to a consistent picture of the 
behaviour of aQCD from below the deconfinement transition to well above
the chiral transition. We have found evidence for a second order chiral
transition at $\beta_c=5.624(2)$, compatible with the $3d$ $O(2)$ 
universality class. The expected $3d$ Goldstone effect in the intermediate
phase $\beta_d< \beta < \beta_c$, which was already noted in Ref.\  
\cite{Karsch:1998qj}, was fully confirmed. The same applies to the first
order deconfinement transition at $\beta_d=5.236(3)$. Our results are
visualized in Fig.\ \ref{3d_chi_fits}, where we show the chiral condensate
data at small mass values together with fits based on our understanding
of the behaviour of aQCD at finite temperature.

%%%%%%%%%%%%%%%%%%%%%%%%%%%%%%%%%%%%%%%%%%%%%%%%%%%%%%%%%%%%%%%%%%%%%%%%%%%%%%%%
\vskip 0.2truecm
\noindent{\Large{\bf Acknowledgments}}

%%%%%%%%%%%%%%%%%%%%%%%%%%%%%%%%%%%%%%%%%%%%%%%%%%%%%%%%%%%%%%%%%%%%%%%%%%%%%%%%
\n We thank Frithjof Karsch and Martin L\"utgemeier for many discussions of 
their work on QCD with adjoint fermions and the use of their programs and data.
We are also grateful to Edwin Laermann for clarifying comments on the chiral
susceptibility and on algorithmic problems. Our work was supported by the
Deutsche Forschungs\-ge\-meinschaft under Grants No.\ FOR 339/2-1 and No.\
En 164/4-4.

\newpage
%%%%%%%%%%%%%%%%%%%%%%%%%%%%%%%%%%%%%%%%%%%%%%%%%%%%%%%%%%%%%%%%%%%%%%%%%%%%%%%%

%%%%%%%%%%%%%%%%%%%%%%%%%%%%%%%%%%%%%%%%%%%%%%%%%%%%%%%%%%%%%%%%%%%%%%%%%%%%%%%%

\end{document}